\documentclass[11pt]{article}
\pdfoutput=1
\usepackage{amstext,amsfonts,amssymb,amscd,amsbsy,amsmath,graphicx,array,float,longtable}
\usepackage[hmargin=1in,vmargin=1.22in]{geometry}
\usepackage[nosort]{cite}
\usepackage{dsfont}
\usepackage{amsfonts}
\usepackage{amssymb}
\usepackage{amsmath}
\usepackage{graphicx}
\usepackage{epstopdf}
\usepackage{slashed}
\usepackage{setspace}
\usepackage{multirow}
\usepackage{booktabs}
\usepackage{color}
\usepackage{mathtools}
\usepackage[
      colorlinks=true,
      linkcolor=darkblue,  
      urlcolor=blue,    
      filecolor=blue,     
      citecolor=red,
linktocpage=true,
      pdfstartview=FitV,
      bookmarksopen=true    
      ]{hyperref}

\definecolor{darkblue}{rgb}{0.2, 0, 0.8}

\newcommand{\UV}{\text{UV}}
\newcommand{\IR}{\text{IR}}
\newcommand{\WZ}{\text{WZ}}

\newcommand{\reef}[1]{(\ref{#1})}

\def\be{\begin{equation}}
\def\ee{\end{equation}}
\def\bea{\begin{eqnarray}}
\def\eea{\end{eqnarray}}

\def\pa{\partial}
\def\m{\mu}
\def\n{\nu}
\def\r{\rho}

\begin{document}
\setstretch{1.05}

\begin{titlepage}

\begin{flushright}
MCTP-12-23\\
\end{flushright}
\vspace{.1cm}


\begin{center}
\bf \Large
 RG flows in $d$ dimensions, the dilaton effective action, 
 \\
 and the $a$-theorem
\end{center}
\vspace{2mm}
\begin{center}
{\bf Henriette Elvang and Timothy M.~Olson}\\
\vspace{0.7cm}
{{\it Randall Laboratory of Physics, Department of Physics,}\\
{\it University of Michigan, Ann Arbor, MI 48109, USA}}\\[4mm]
{\small \tt  elvang@umich.edu, timolson@umich.edu}
\end{center}
\vspace{-2mm}

\begin{abstract}
Motivated by the recent dilaton-based proof of the 4d $a$-theorem, we study the dilaton effective action for RG flows in $d$ dimensions. When $d$ is even, the action consists of a Wess-Zumino (WZ) term, whose Weyl-variation encodes the trace-anomaly, plus all Weyl-invariants. For $d$ odd, the action consists of Weyl-invariants only.  We present explicit results for the flat-space limit of the dilaton effective action in $d$-dimensions up to and including 8-derivative terms. GJMS-operators from conformal geometry motivate a form of the action that unifies the Weyl-invariants and anomaly-terms into a compact general-$d$ structure. 

A new feature in 8d is the presence of an 8-derivative Weyl-invariant that pollutes the O($p^8$)-contribution from the WZ action to the dilaton scattering amplitudes; this may challenge a dilaton-based proof of an $a$-theorem in 8d. 

We use the example of a free massive scalar for two purposes: 1) it allows us to confirm  the structure of the $d$-dimensional dilaton effective action explicitly; we carry out this check for $d=3,4,5,\dots,10$; and 2) in 8d we demonstrate how the flow $\Delta{a} = a_\text{UV} - a_\text{IR}$ can be extracted systematically from the O($p^8$)-amplitudes despite the contamination from the 8-derivative Weyl-invariant. This computation gives a value for the $a$-anomaly of the 8d free conformal scalar that is shown to match the value obtained from  zeta-function regularization of the log-term in the free energy.

\end{abstract}

\end{titlepage}

\vspace{4mm}
\setcounter{tocdepth}{2}
\tableofcontents

\newpage

\setcounter{equation}{0}
\section{Introduction \& Summary}

The recent proof of the 4d $a$-theorem by Komargodski and Schwimmer (KS) \cite{KS} makes exquisite use of the low-energy effective interactions of the dilaton, a field that can be thought of as the Goldstone mode of spontaneously broken conformal symmetry or as  a compensator background field. KS showed, following earlier work \cite{spon}, that the form of the dilaton effective action is dictated by its Weyl-transformation properties and that the low-energy behavior of the 4-point dilaton scattering amplitude 
\be
  \mathcal{A}_4(s,t) = \Delta{a}\, \frac{4}{f^4} (s^2 + t^2 + u^2)
  \label{zohar4pt}
\ee
encodes the flow of the $a$-anomaly $\Delta{a}=a_\text{UV}-a_\text{IR}$. In the forward limit $t \to 0$,  $A_4(s,0)/s^{3}$ has a simple pole at $s=0$ whose residue is $4\Delta{a}/f^4$. A  contour integral argument then gives
\bea
  \Delta{a} 
  ~=~ \frac{f^4}{4\pi} \int_{0}^\infty ds\, \frac{{\rm Im}\,A(s,0)}{s^3}  
  ~>~ 0 \,.
\eea 
Since the integrand on the RHS is positive definite, this proves $a_\text{IR} < a_\text{UV}$ for an RG flow from a 4d UV CFT to a 4d IR CFT. The convergence of the dispersion integral has been clarified in \cite{K,joep2}.

Zamolodchikov's $c$-theorem \cite{zamo} and the Cardy-KS $a$-theorem \cite{cardy,KS} demonstrate the irreversibility of the RG flow between 2d and 4d CFT fixed points, respectively. It is interesting to ask if this property  generalizes to other dimensions. Holographic arguments indicate that it does generalize, and they provide an interesting connection to entanglement entropy \cite{Myers:2010xs,Myers:2010tj,casini9}. In even dimensions, the irreversibility of the flow can be encoded in an $a$-theorem for the `type A' anomaly $a$ associated with the Euler density term in the trace anomaly polynomial \cite{Deser:1993yx}\footnote{We normalize the $d=2k$-dimensional Euler density as
\bea
\label{eulerdens}
E_{2k}(g_{\mu\nu}) &=& 
\frac{1}{2^k}\,R_{\mu_1\nu_1}{}^{\rho_1\sigma_1} \dots R_{\mu_k\nu_k}{}^{\rho_k\sigma_k}  
\,\epsilon_{\rho_1\sigma_1\dots\rho_k\sigma_k} 
\,\epsilon^{\mu_1\nu_1\dots\mu_k\nu_k}  \,.
\eea
The `type B' anomalies $c_i$ multiply a set of independent Weyl-invariant scalars $\sqrt{-g} I_i$; there is 1 in 4d, 3 in 6d, and \cite{Boulanger:2004zf} found 12 in 8d. The $c$-anomalies do not always decrease along an RG flow.}
\be
\label{traceT}
\left\langle T_\mu{}^\mu\right\rangle = \sum\limits_i c_i I_i - (-)^{d/2}a\, E_d
\,.
\ee
In odd dimensions, the constant term in the free energy $F= -\log Z$ offers a candidate for an analogous $F$-theorem; for recent work see \cite{fthem,Klebanov:2011gs}.

The dilaton can be introduced in even as well as in odd dimensions, and one may ask what information can be extracted from its low-energy effective action: in particular if it can be used to prove a higher-$d$ $a$-theorem and whether it plays a  role for odd-$d$ RG flows. The focus of this paper is to study the structure of the dilaton effective action in general $d$ dimensions.

The dilaton-based approach \cite{KS,K} to the $a$-theorem was examined recently in \cite{6d} for RG flows between 6-dimensional CFTs. The 6d dilaton effective action was constructed up to 6-derivative order and its structure verified in explicit examples. The examples  also served to clarify the distinctive roles of the dilaton in the cases of spontaneous and explicit breaking of the conformal symmetry. In the former case, the dilaton is a dynamical field of the low-energy theory and it contributes as such to the scattering amplitudes via Feynman diagrams with internal dilaton lines; this was demonstrated explicitly with the example of the 6d (2,0) theory on the Coulomb branch \cite{6d}. (See also \cite{Maxfield:2012aw}.) On the other hand, when the conformal symmetry is  broken explicitly,  the dilaton is introduced as a compensator field which can be made arbitrarily weakly coupled such that in the low-energy scattering amplitudes it may be treated as a source field. This case was illustrated in \cite{6d} by the example of the 6d free massive scalar.

The simple KS approach to proving $a_\text{IR} < a_\text{UV}$ in 4d does not directly carry over to 6d \cite{6d}, and no general proof of the 6d $a$-`theorem' has yet been offered. One technical difficulty with generalizing the KS dispersion relation argument is that in 6d the anomaly flow $\Delta{a}$ is associated with the 6-derivative terms in the action: the 4-dilaton amplitude is then $A_4(s,t) \sim \Delta{a}\,s t u$, and since it vanishes in the forward limit no clean positivity statement is extracted; for details see \cite{6d}.

It would seem easier to derive a positivity result based on a 4-point amplitude of the form $A_4(s,t) \sim (s^4 + t^4 +u^4)$. Indeed in 8d, the $a$-anomaly is associated with 8-derivative terms in the action, and at order $O(p^8)$ the 4-dilaton amplitude takes this form. However, the 8d dilaton effective action also contains an 8-derivative Weyl-invariant that contributes non-trivially to the scattering amplitudes \cite{6d}. In fact, proving positivity of the coefficient of the $O(p^8)$-terms in the 4-dilaton amplitude amounts to proving only that the coefficient of this new Weyl-invariant is positive and does not yield $\Delta{a} >0$. 

One purpose of this paper is to clarify the structure of the terms in the dilaton effective action in 8d up to and including 8-derivative terms. We will also show that despite the pollution from the 8-derivative Weyl-invariant, the flow $\Delta{a}$ can be extracted systematically. We demonstrate this explicitly for the example of the 8d free massive scalar. The result for $\Delta{a} = a_\text{scalar,8d}$ agrees with that found using zeta-function regularization of the coefficient of the log-term in the free energy. 

It must be noted that the study of 8d RG flows is motivated by the wish to understand the structure of the dilaton effective action and the generality of the dilaton-based approach of KS in even dimensions. We know of no examples of interacting 8-dimensional conformal theories (and there can be no superconformal ones \cite{nahm}), so an 8d (or higher-$d$) $a$-theorem may be of limited applicability. 

The analysis in 8d is part of our more general study of the dilaton effective action in $d$ dimensions, with $d$  even or odd. The trace-anomaly exists only for even $d$, and therefore it is \emph{a priori} clear that for $d$ odd, the low-energy dilaton effective action simply consists of a derivative-expansion of Weyl-invariants.\footnote{A holographic approach to the dilaton effective action in $d$-dimensions was recently discussed in \cite{janetandherfriends}.} Such Weyl-invariants must also be included when writing down the dilaton effective action in even $d$,  in addition to the Wess-Zumino action whose Weyl-variation produces the integral of the trace anomaly polynomial \reef{traceT}.

Despite the obvious difference between even and odd $d$, we find a compact unifying form for the terms in the dilaton effective action that contribute with non-vanishing local matrix elements to the on-shell dilaton amplitudes; these are the terms that are non-vanishing under the equations of motion. For flows induced by explicit breaking of  conformal symmetry this is all that is needed.
The unified form is given in terms of 
\bea 
\label{Wkflat}
   \mathcal{W}_k ~=~\big(\tfrac{2}{d-2k}\big)^2 \,e^{-(d/2-k)\tau} P_k \,  e^{-(d/2-k)\tau} \, ,
\eea 
where $\tau$ is the dilaton field and $P_k = (\Box^k + \text{curvature terms})$ is a GJMS-operator \cite{GJMS}. The GJMS operators $P_k$ are higher-derivative generalizations of the conformal Laplacian ($k=1$) and the Paneitz operator ($k=2$) \cite{paneitz}.\footnote{Although commonly referred to as the Paneitz operator in the math literature, this 4-derivative operator actually first appeared in Fradkin and Tseytlin's work \cite{FradkinTseytlin} from 1982.} The `covariance' of $P_k$ under conformal transformations (see \reef{gjmsOP}) ensures that $\mathcal{W}_k$ behaves well under Weyl-transformations, $\tau \to\tau +\sigma$ and $g_{\mu\nu} \to e^{2\sigma}g_{\mu\nu}$: for $k \ne d/2$ it transforms as
$\mathcal{W}_k \to  e^{-d\,\sigma} \mathcal{W}_k$ so that $\sqrt{-g}\mathcal{W}_k$ is Weyl-invariant. We find that up to and including 8-derivative terms, the $d$-dimensional action in flat space can be written
\be
   S
   = \int d^dx \,\sqrt{-g} 
   \Big[
   \tfrac{(d-2)^2}{8} f^{d-2}\,\mathcal{W}_1
   + \alpha^{(d)} \, \mathcal{W}_2
   + \beta^{(d)} \, \mathcal{W}_3
   + \gamma^{(d)} \,\mathcal{W}_4 
   + \tilde{\gamma}^{(d)} \, \tfrac{(d-4)^2}{4} \,e^{d\,\tau}\,(\mathcal{W}_2)^2
   + \dots
   \Big]\bigg|_{g_{\mu\nu}=\eta_{\mu\nu}}\!,
   \label{Sgjms-intro}
\ee 
where the ellipses stand for terms that vanish on-shell. If $d$ is even, $\sqrt{-g}\mathcal{W}_{d/2}$ reduces to $\tau \Box^{d/2} \tau$ in flat space. This is not Weyl-invariant, and it is known from $d=4$ \cite{K} and $d=6$ \cite{6d} that this form encodes\footnote{The $d$-dimensional results can be obtained from a generalization of the analysis in section 2 of \cite{K}; or it can be motivated by an argument  \cite{JanetandRob} based on Branson's $Q$-curvature \cite{branson}.} the anomaly flow as $\tfrac{d}{2}\Delta{a} \,\tau \Box^{d/2}\tau$. We demonstrate it in this paper for $d=8$. So in $d=4,6,8$ one simply re-interprets the coefficient of $\mathcal{W}_{d/2}$ as $\tfrac{d}{2}\Delta{a}$.

The action \reef{Sgjms-intro} should be thought of as a generator of the dilaton amplitudes for the case of flows induced by explicit
breaking of conformal symmetry. In general backgrounds, the
GJMS-operators $P_k$ exist for all $k$ for $d$ odd, but only for $k \le d/2$ when $d$ is even \cite{GJMS}; $k=d/2$ is of course the order where the trace anomaly enters. However, for conformally flat backgrounds, the GJMS-operators exist for all $k$ in both even and odd dimensions \cite{Juhl}.

We carry out a non-trivial test of the result for the dilaton effective action \reef{Sgjms-intro} using the example of the $d$-dimensional free massive scalar. In this example, the dilaton is introduced as a compensator to restore conformal symmetry. The massive scalar couples quadratically to the dilaton, so the $n$-dilaton amplitudes can be calculated as 1-loop amplitudes with the massive scalar running in the loop. The low-energy expansion of these 1-loop amplitudes can then be compared with the dilaton amplitudes produced by $S$ in \reef{Sgjms-intro}. We obtain a perfect match; the specific coefficients $\alpha^{(d)}$, $\beta^{(d)}$, $\gamma^{(d)}$ and $\tilde{\gamma}^{(d)}$ of the action \reef{Sgjms-intro} are listed in table \ref{table1} in section \ref{s:scalarDd} for $d=3,4,5,\dots, 10$. 

We discuss the structure of the action \reef{Sgjms-intro} further in sections \ref{s:gjms} and \ref{s:higher}; and we show that at the order of 10- and 12-derivatives, the GJMS-based building blocks $\mathcal{W}_k$ are not sufficient and new structures are needed. Perhaps this points to possible generalizations of the GJMS-operators. 

The paper is structured as follows. In section \ref{s:action} we analyze the dilaton effective action in $d$ dimensions order by order in derivatives up to $O(\pa^8)$ and calculate the corresponding dilaton matrix elements, assuming the context of explicit breaking and hence an arbitrarily weakly coupled dilaton. 
In section \ref{s:scalar8d} we study the example of the free massive scalar in 8d and show how to systematically extract $\Delta{a}$ from the dilaton amplitudes. We review in section \ref{s:anomd} how the $d$-dimensional anomaly can be calculated as the coefficient of the log-term in the free energy for the free massive scalar and explicitly verify a compact formula for $a_\text{scalar}$ by Diaz \cite{Diaz:2008hy} for $d=4,6,\dots, 20$. In particular, the $d=8$ result matches that of our dilaton amplitude calculation in section \ref{s:scalar8d}. We generalize the analysis of the free massive scalar to $d$-dimensions in section \ref{s:scalarDd} and use it to verify the general result for the dilaton effective action. Details of our calculations can be found in four appendices.

\setcounter{equation}{0}
\section{Dilaton effective action and scattering in $d$ dimensions}
\label{s:action}
The dilaton effective action $S$ consists of diff$\times$Weyl invariant terms and in even dimensions the Wess-Zumino action whose Weyl variation produces the trace anomaly,
\be
   \delta_\sigma S_\WZ = \int d^dx 
    \sqrt{-g}\,\sigma \left\langle T_\mu{}^\mu\right\rangle = 
   \int d^dx \sqrt{-g}\,\sigma \Big( \sum\limits_i c_i I_i - (-)^{d/2}a\, E_d \Big)\,.
\ee
The construction of $S_\WZ$ was detailed in \cite{KS,spon,6d} and results given explicitly for $d=4,6$; we outline the construction for $d=8$ in appendix \ref{app:euler8d} and discuss the result in section \ref{s:S8deriv}.

In a spacetime with fixed background metric $g_{\mu\nu}$, the diff$\times$Weyl invariant terms are curvature scalars constructed from the `hatted' metric $\hat{g}_{\mu\nu}=e^{-2\tau}g_{\mu\nu}$, where $\tau$ is the dilaton field. Here we are concerned with the dilaton effective action in flat space, so in the following we take
\be
  \hat{g}_{\mu\nu}=e^{-2\tau} \eta_{\mu\nu} \,.
\ee
For a conformally flat background, any appearance of the Riemann tensor can be replaced by the Weyl tensor plus Ricci scalar and tensor terms via \reef{RiemannWeyl}. Thus we can construct our Weyl-invariants  from  the Ricci scalar, Ricci tensor and covariant derivatives thereof.
Examples are $\hat{R}_{\mu\nu}\hat{R}^{\mu\nu}$ and $\hat{R} \Box \hat{R}$.

We organize the dilaton low-energy effective action as a derivative expansion
\bea
  \label{SdilatonGen}
  S&= &
  \underbrace{~S^{\pa^2}~}_{\text{\reef{S2deriv}}} 
  ~+~  \underbrace{~S^{\pa^4}~}_{\text{\reef{S4deriv}}} 
  ~+~  \underbrace{~S^{\pa^6}~}_{\text{\reef{S6deriv}}} 
  ~+~  \underbrace{~S^{\pa^8}~}_{\text{\reef{S8deriv}}} 
  + ~\dots\\
  \nonumber
  \text{\scriptsize compact form:} \!\!\!\!
  && \hspace{1.5cm}~~{\text{\scriptsize \reef{S4deriv2}}}
  ~~~~~~\,~{\text{\scriptsize \reef{S6derivGJMS}}}
  ~~~~~~~{\text{\scriptsize \reef{p8gjms}}}
\eea
$S_\WZ$ is included as part of the $d$-derivative action for $d$ even. 
In the following, we systematically construct $S^{\pa^{2k}}$ for $k=1,2,3,4$ in $d$-dimensions.\footnote{All tensor manipulations were done through a combination of pencil, paper, and the Mathematica package xAct \cite{xact}.}
The equation reference given below each term in \reef{SdilatonGen} indicates  where to find the result at order $O(\pa^{2k})$. The compact form refers to the terms in GJMS-form \reef{Sgjms-intro} discussed in the Introduction. 
Before we analyze each $S^{(\pa^{2k})}$ and calculate the $O(p^{2k})$ scattering amplitudes, let us make a few general comments:
\begin{itemize}
\item
{\bf \em Physical dilaton.}
In order to calculate dilaton scattering amplitudes, we introduce the physical dilaton field $\varphi$ defined by 
\begin{align}
 \label{physdilaton}
e^{-\frac{d-2}{2}\,\tau} ~=~ 
  1 - \frac{\varphi}{f^{(d-2)/2}} 
  ~=~ \Omega\, f^{-(d-2)/2}  \,.
\end{align}
This definition ensures that the physical dilaton has on-shell condition $k^2=0$.

\item
{\bf \em Explicitly broken conformal symmetry.}
In this paper, we focus entirely on the scenario of explicitly broken conformal symmetry. This means that we treat the dilaton as arbitrarily weakly coupled, so that any contributions to the dilaton amplitudes from diagrams with internal dilaton lines are suppressed \cite{6d}. As a result, the low-energy dilaton amplitudes at $O(p^{2k})$ derive solely from the contact-terms with $2k$ derivatives. The only terms in the action that contribute to the amplitudes are therefore those that do not vanish on the leading order (i.e. 2-derivative) dilaton equations of motion. 

\item 
{\bf \em From action to amplitudes.}
In the dilaton effective action we find terms such as
\begin{align}
\label{gjms}
e^{-\frac{d-2k}{2}\,\tau}\, \square^k e^{-\frac{d-2k}{2}\,\tau}\,.
\end{align}
Expanding \reef{gjms} in powers of $\varphi$ gives terms $\varphi^{n_2} \square^k \varphi^{n_1}$ whose contributions to the $n$-point matrix elements are easy to compute:
\begin{align}
\varphi^{n_2} \square^k \varphi^{n_1} 
~~\rightarrow ~~
n_1! \,n_2! \sum\limits_{1 \le i_1<i_2<\dots <i_{n_1} \le n}s_{i_1\,i_2\,\dots \, i_{n_1}}^k\,.
\label{gjmspolys}
\end{align}
Here $n=n_1+n_2$ and the Mandelstam invariants are defined as
\be
 \label{sijs}
s_{i_1\,i_2\,\dots\, i_\ell} = - (p_{i_1}+p_{i_2}+\dots p_{i_\ell})^2 \,.
\ee
The simplicity of \reef{gjmspolys} was exploited previously in  \cite{6d} for the  calculation of the dilaton matrix elements in 4d and 6d. In this paper, the action also contains terms such as $\varphi\square^2\varphi^2\square^2\varphi^2$ which produce polynomials of the form $(s_{12}^2s_{34}^2+\text{perms})$.
\end{itemize}

\subsection{Kinetic term and dilaton equations of motion}
The kinetic term for the dilaton is generated by the unique 2-derivative diff$\times$Weyl invariant   $\sqrt{-\hat{g}}\hat{R}$. In a flat background we have
\bea
   \label{ricciS}
   \sqrt{-\hat{g}}\,\hat{R}
   ~=~
   2(d-1)\Big( \Box \tau - \frac{d-2}{2} (\pa\tau)^2 \Big)\, e^{-(d-2)\tau} \, ,
\eea
so  after partial integration we can write 
\bea
S^{\pa^2} \,=\,
-\tfrac{1}{8}\tfrac{d-2}{d-1}\,f^{d-2}\, \int d^d x\sqrt{-\hat{g}}\,\hat{R} 
\,=\,
 -\tfrac{(d-2)^2}{8}\,f^{d-2}\,\int d^d x \,(\partial \tau)^2\, e^{-(d-2)\tau}
\,=\,
-\tfrac{1}{2} \,\int d^d x \,(\partial \varphi)^2
 \,.~~
 \label{S2deriv}
\eea
The constant $f$ has dimension of mass, and the overall factor is chosen such that the physical dilaton, $\varphi$ defined in \reef{physdilaton}, has a canonically normalized kinetic term.

It follows from \reef{S2deriv} that the dilaton equation of motion is
\begin{align}
\label{eom}
\square\tau = \frac{d-2}{2}(\partial\tau)^2 
\text{\hspace{0.5cm} or \hspace{0.5cm}} 
\square\varphi = 0 \,.
\end{align}
The latter form tells us that the on-shell condition for the physical dilaton is $k^2=0$, as noted below \reef{physdilaton}. Note that by \reef{ricciS}, $\hat{R}$ vanishes on-shell.

\subsection{4-derivative action}

There are two  4-derivative Weyl-invariants, $\sqrt{-\hat{g}}\,\hat{R}^2$\, and\, $\sqrt{-\hat{g}}\,\big(\hat{R}_{\mu\nu}\big)^2 \equiv \sqrt{-\hat{g}}\,\hat{R}_{\mu\nu}\hat{R}^{\mu\nu}$, so we write the  4-derivative action
\begin{align}
\label{S4deriv}
S^{\partial^4} = \int d^d x\sqrt{-\hat{g}}\left[ \alpha_1 \hat{R}^2 + \alpha_2 \left(\hat{R}_{\mu\nu}\right)^2\right] - \delta_{d,4} \,\Delta a \,S_{\WZ}\,,
\end{align}
where $\alpha_i$ are constants. 
In 4d, the flat-space limit of $S_{\WZ}$ is \cite{KS,spon}
\be  \label{4dsanom}
S_\WZ
= \int d^4x \, \Big[-4(\pa \tau)^2\Box\tau + 2 (\pa\tau)^4\Big]\,.
\ee
The Weyl-invariant  $\hat{R}^2$ is zero on-shell, but 
\bea
\nonumber
\sqrt{-\hat{g}}\,\big(\hat{R}_{\mu\nu}\big)^2
\hspace{-3mm}
&\!\!=&
  \hspace{-3mm}
 \frac{(d-2)^2}{2} 
 \bigg[ \frac{2d(d-1)}{(d-2)^2} (\square \tau )^2
 -\frac{3 d^2-8 d+8}{(d-2)} (\square \tau ) (\partial \tau )^2 
	+\big(d^2-4 d+6\big) (\partial \tau )^4
	\bigg]\,e^{-(d-4) \tau}\\[2mm]
&\!\!\!\!\xrightarrow{\text{EOM}}\!\!&	
- \frac{1}{4}  (d-4) (d-2)^2 \,(\partial \tau )^4\,e^{-(d-4) \tau} \,.
\label{ricciSQ}
\eea
This vanishes  in $d=4$, as found in \cite{KS}, so the WZ term is the only contribution to the $O(p^4)$ matrix elements in 4d. This feature facilitated KS's proof of the $a$-theorem. For $d \ne 4$, the Weyl-invariant  \reef{ricciSQ} gives non-vanishing contributions to the $O(p^4)$ matrix elements, as in 6d \cite{6d}.

The 4-derivative action \reef{S4deriv} can be written compactly as 
\begin{align}
\label{S4deriv2}
S^{\partial^4} = \int d^d x\bigg[ \alpha \,\bigg(\frac{2}{d-4}\bigg)^2 
  e^{-\frac{d-4}{2}\,\tau} \,\square^2 e^{-\frac{d-4}{2}\,\tau}+\dots\bigg] \, ,
\end{align}
where the ``$\dots$'' refer to terms that vanish on-shell.  To see how \reef{S4deriv2} can be compatible with the WZ term, note:
\begin{itemize}
\item For $d \ne 4$, the WZ term is absent, and straightforward algebra with the expressions in \reef{4derivAppB} shows that 
$e^{-\frac{d-4}{2}\,\tau} \,\square^2 e^{-\frac{d-4}{2}\,\tau} =\tfrac{4-d}{(d-2)^2} \sqrt{-\hat{g}}\,\big(\hat{R}_{\mu\nu}\big)^2 + $ terms that vanish on-shell. 
\item In $d=4$, the only contributions to the $O(p^4)$ dilaton matrix elements come from the WZ action. As noted in \cite{K}, the flat-space limit of the 
$S_{\WZ}$ can be written in terms of $-2 \Delta a \, \tau \Box^2 \tau+ $ terms that vanish on-shell. But this is exactly the expression recovered in the limit $d \to 4$ of \reef{S4deriv2} with $\alpha=2 \Delta a$.
\end{itemize}
Practically both cases above require solving a system of 3 equations, matching the coefficients of each unique type of term in \reef{4derivAppB} --- $(\square\tau)^2$, $(\partial\tau)^2\square\tau$, and $(\partial\tau)^4$ --- with \reef{S4deriv2}, using only the 2 variables $\alpha_1$ and $\alpha_2$ \cite{K}. It is noteworthy that a solution exists.

The $n$-point matrix elements at order $O(p^4)$ can be expressed in terms of a basis of  polynomials in the Mandelstam invariants \reef{sijs} as 
\be
\label{basisPp4}
  P_n^{(4)} ~\equiv \sum_{1\le i<j\le n} s_{ij}^2 \,.
\ee

As an example, consider the $O(p^4)$ to the amplitudes determined by $S^{\pa^4}$ in \reef{S4deriv2}. Changing variables from $\tau$ to $\varphi$ via \reef{physdilaton} gives
\bea
  \alpha\, \bigg( \frac{2}{(d-2)^2}\bigg)^2 
  \frac{1}{f^{2d-4}} \, \varphi^2 \Box^2 \varphi^2 
  ~+~ O(\varphi^5) \,,
\eea
so  using \reef{gjmspolys} we can directly read off the 4-point amplitude 
\bea
 \label{A44}
 \mathcal{A}_4^{(4)}=
 \alpha\, \bigg( \frac{2}{(d-2)^2}\bigg)^2 
  \frac{1}{f^{2d-4}} \, 2! \,2!  
  \sum_{1\le i<j \le 4} s_{ij}^2
  ~=~
   \alpha\, \frac{32}{(d-2)^4} \frac{1}{f^{2d-4}} \big( s^2 + t^2 + u^2 \big) \,.
\eea 
For $d=4$, we identify $\alpha = 2 \Delta a$, so \reef{A44} agrees with the result \reef{zohar4pt}.
Taking $d=6$, we find 
$\mathcal{A}_4^{(4)} =  \frac{\alpha}{8 f^8} \big( s^2 + t^2 + u^2 \big)$. This matches the result in eq.~(3.18) of \cite{6d} in which $\mathcal{A}_4^{(4)} $ was expressed in terms of a coefficient $b$ related to $\alpha$ by  $\alpha =4b$. 

The higher-point matrix elements of $S^{\pa^4}$ are straightforward to extract from \reef{S4deriv2}. To avoid cluttering the main text, we list the $O(p^4)$ matrix elements in \reef{ADimd_p4} for the general $d$-dimensional case. 

\subsection{6-derivative action}

For a $d$-dimensional conformally flat metric, any 6-derivative Weyl-invariant can be written in terms of
\be
\hat{R}^3\,, ~~~~~~
\hat{R}\big(\hat{R}_{\m\n}\big)^2\equiv 
\hat{R}\big(\hat{R}^{\m}{}_{\n}\hat{R}^{\n}{}_{\m}\big)\,, ~~~~~~
\hat{R}\,\hat{\square}\hat{R}\,, ~~~~~~
\text{and}~~~~~~
\big(\hat{R}_{\m\n}\big)^3 
\equiv 
\big(\hat{R}^{\m}{}_{\n}\hat{R}^{\n}{}_{\r}\hat{R}^{\r}{}_{\m}\big), 
\label{6derivBasis}
\ee
up to total derivatives. 
For example \cite{anselmi3}, 
\begin{align}
\hat{R}_{\m\n}\hat{\square}\hat{R^{\m\n}}&=
  \frac{1}{(d-2) (d-1)}\hat{R}^3
  -\frac{2d -1}{(d-2) (d-1)}\hat{R}\big(\hat{R}_{\m\n}\big)^2\nonumber\\
	&\qquad+\frac{d}{4 (d-1)}\hat{R}\hat{\square}\hat{R}
	+\frac{d}{d-2}\big(\hat{R}_{\m\n}\big)^3 
	+ \text{total derivatives}\,.
\end{align}
Using the basis \reef{6derivBasis}, the 6-derivative action takes the form:
\begin{equation}
S^{\partial^6} = \int d^d x\sqrt{-\hat{g}}\left[ \beta_1\, \hat{R}^3 + \beta_2\, \hat{R}\big(\hat{R}_{\m\n}\big)^2 + \beta_3\, \hat{R}\hat{\square}\hat{R} + \beta_4 \,\big(\hat{R}_{\m\n}\big)^3\right] + \delta_{d,6}\,\Delta a \, S_{\WZ}\,,
 \label{S6deriv}
\end{equation}
where $\beta_i$ are constants.\footnote{The observant reader may notice that the 6-derivative action for $d=6$ in \cite{6d} contains only three curvature invariants. This is sufficient in 6d because one can use that the Euler density $E_6$ is a total derivative to eliminate one of the four invariants.} A curved-space derivation of the WZ action in $d=6$ dimensions is given in \cite{6d} and  \cite{Maxfield:2012aw}. The flat space limit is
 \be \label{euler6}
  S_\WZ
  =
  \int d^6 x\, \Big[
  -24 (\Box\tau)^2 (\pa\tau)^2 + 24 (\pa\tau)^2  (\pa\pa\tau)^2
  + 36 \Box\tau  (\pa\tau)^4
  - 24 (\pa\tau)^6
  \Big] \,.
 \ee
Explicit expressions for each of the four Weyl-invariants are available in appendix \ref{app:WeylInvariants}. The three invariants proportional to $\hat{R}$ vanish on-shell, so only 
$\big(\hat{R}_{\m\n}\big)^3$ contributes to the $O(p^6)$ dilaton matrix elements:
\begin{align}
\big(\hat{R}_{\m\n}\big)^3 &~\xrightarrow{\text{EOM}} ~
- \frac{1}{4} e^{-(d-6) \tau } (d-6) (d-2)^3 
\Big((\partial \partial \tau )^2 (\partial \tau )^2-2 (\partial \tau )^6\Big)\,.
\end{align}
This  vanishes in $d=6$, hence for the case of explicitly broken conformal symmetry, only  the WZ action generates contact-term contributions to the 6d matrix elements at $O(p^6)$.\footnote{In the case of spontaneous breaking, the 4-derivative terms also contribute through pole diagrams \cite{6d}.}

The 6-derivative action \reef{S6deriv} can be written in the  compact form
\begin{align}
\label{S6derivGJMS}
S^{\partial^6} 
= \int d^d x \left[\,\beta\left(\frac{2}{d-6}\right)^2 
e^{-\frac{d-6}{2}\,\tau}\, \square^3 e^{-\frac{d-6}{2}\,\tau}+ \dots\right],
\end{align}
where the ``$\dots$'' refer to terms that vanish on-shell. When $d \to 6$, the action \reef{S6derivGJMS} reduces to $3 \Delta a \,\tau \Box^3 \tau$ and we identify $\beta = 3\Delta a$. 
The equivalence of \reef{S6deriv} and \reef{S6derivGJMS} requires a solution to  an overconstrained system of 8 equations (from matching the coefficients of the 8 distinct terms in the expressions of \reef{6derivAppBa}-\reef{6derivAppBd}, e.g. $(\square\tau)(\square^2\tau)$, with \reef{S6derivGJMS}) using only the 4 variables $\beta_1,\dots,\beta_4$ from \reef{S6deriv}.

The local matrix elements with $n\le 8$ external dilatons can at  $O(p^6)$ be expressed in terms of two linearly independent symmetric Mandelstam polynomials,
\bea
\label{basisPp6}
 P^{(6)}_{n,A} ~= \sum\limits_{1\leq i<j\leq n}s_{ij}^3
 \,, \hspace{8mm} 
 P^{(6)}_{n,B}~ =  \sum\limits_{1\leq i<j<k\leq n}s_{ijk}^3\,.
\eea
Writing the amplitudes in this basis requires identities such as
\begin{align}
\label{polyRel}
\sum\limits_{1\leq i<j<k<l\leq8} \!\!s_{ijkl}^3 ~
= ~-2 P^{(6)}_{8,A}~ + ~2P^{(6)}_{8,B} \,.
\end{align}

We list the $O(p^6)$ amplitudes in \reef{ADimd_p6}. In 6d, the 4-, 5- and 6-point amplitudes reproduce (3.19)-(3.21) of \cite{6d} with $\beta=3\Delta a$. For example,
\be
\mathcal{A}_6^{(6)}~=~ \frac{64 (d+2)}{(d-2)^6} \frac{\beta}{f^{3 d-6}}\left(4 \,d \,P_{6,A}^{(6)} + (d+2)P_{6,B}^{(6)}\right)
~~\xrightarrow{d\to6\,}~~
  \frac{3\Delta a}{f^{12}}\left(3 \,P_{6,A}^{(6)} + P_{6,B}^{(6)}\right)\,.~~
\ee
In 8d, we have 
\be
\mathcal{A}_6^{(6)} 
~~\xrightarrow{d\to8\,}~~
\frac{20 \beta}{729 f^{18}}\left(16 \,P_{6,A}^{(6)} + 5 P_{6,B}^{(6)}\right)\,.~~
\label{A6p6in8d}
\ee
We  match this and the other $O(p^6)$ $n$-point amplitudes, $n\le 8$, for the example of the free massive scalar in sections \ref{s:scalar8d} and \ref{s:scalarDd}.

\subsection{8-derivative action}
\label{s:S8deriv}
For a $d$-dimensional, conformally flat metric, we find nine independent Weyl-invariants (up to total derivatives) by explicit calculation, so the off-shell action can be written as
\begin{align}
S^{\partial^8} &= \int d^d x\sqrt{-\hat{g}}
\left[ 
\gamma_1\hat{R}^4 
+ \gamma_2\hat{R}^2\big(\hat{R}_{\m\n}\big)^2 
+ \gamma_3\hat{R}\big(\hat{R}_{\m\n}\big)^3 
+ \gamma_4\big((\hat{R}_{\m\n})^2\big)^2 
+ \gamma_5\big(\hat{R}_{\m\n}\big)^4 
+ \gamma_6\big(\hat{\square}\hat{R}\big)^2\right.\nonumber\\
	&\left.
	\hspace{2.8cm}
+ \gamma_7\big(\hat{\square}\hat{R}_{\m\n}\big)^2 
+ \gamma_8\hat{R}\big(\hat{\nabla}_\m\hat{R}\big)^2 
+ \gamma_9\big(\hat{R}_{\m\n}\big)^2\hat{\square}\hat{R}\right] 
~-~ \delta_{d,8}\,\Delta a \,S_{\WZ} \, ,
\label{S8deriv}
\end{align}
with constants $\gamma_i$. We have abbreviated some index contractions, e.g.~$(\hat{R}_{\m\n}\big)^4 
\equiv (\hat{R}^\m{}_{\n}\hat{R}^\n{}_{\r}\hat{R}^\r{}_{\lambda} \hat{R}^\lambda{}_{\m}\big)$, see also \reef{contractions}.

In flat space, the 8d WZ action is 
\bea
\label{SWZ8d}
S_\WZ &=&48 \int d^8  x  \Big[
  3 (\square ^2\tau ) (\partial \tau )^4
  +6 (\square \tau )^3 (\partial \tau )^2
  +36 (\square \tau )^2 (\partial \partial \tau \partial \tau \partial \tau )
  +16 (\square \tau ) (\partial \partial \partial \tau \partial \tau \partial \tau \partial \tau )
\nonumber\\
&& \hspace{2cm}
	-12 (\square \tau ) (\partial \partial \tau )^2 (\partial \tau )^2
	-24 (\partial \partial \tau \partial \tau \partial \tau ) (\partial \partial \tau )^2
	\nonumber\\
	&& \hspace{2cm}
	+12 (\square \tau )^2 (\partial \tau )^4
	-12 (\partial \partial \tau )^2 (\partial \tau )^4
	-20 (\square \tau ) (\partial \tau )^6
	+15 (\partial \tau )^8 \Big]\,.
\eea
Details of the derivation are described in appendix \ref{app:euler8d}.
Applying the equations of motion, we find
\be
S_\WZ~\xrightarrow{\text{EOM}}~
144 \int d^8 x \Big[
-8 (\partial \partial \tau \partial \tau \partial \tau ) \left(\partial \partial \tau )^2\right.
-32 (\partial \partial \tau \partial \tau \partial \tau )^2
-2 \left(\partial \partial \tau )^2\right. (\partial \tau )^4
+3 (\partial \tau )^8\Big]\,.~ \label{Swz8}
\ee
It is clear from \reef{SWZ8d} and \reef{Swz8} that $S_\WZ$ contributes to 5- and higher-point amplitudes, but not to the 4-point amplitude.

We have obtained explicit expressions for the $d$-dimensional Weyl-invariants in \reef{S8deriv}; the procedure for calculating them is straightforward, though to simplify them requires some effort with multiple applications of partial integration. Since the general-$d$ results are rather involved, we present them in appendix \ref{app:WeylInvariants} only for $d=8$.

Six of the nine Weyl-invariants in \reef{S8deriv} vanish on-shell; the only non-vanishing ones are  
$\big((\hat{R}_{\m\n})^2\big)^2$, $\big(\hat{R}_{\m\n}\big)^4$, and 
$\big(\hat{\square}\hat{R}_{\m\n}\big)^2$. These three are also related on-shell; for $d>2$:
\be
 \sqrt{-\hat{g}} \big(\hat{\Box} \hat{R}_{\mu\nu}\big)^2
 ~\xleftrightarrow{\text{\,~EOM~\,}}~\frac{d}{(d-2)^2}
 \Big(
   - \sqrt{-\hat{g}}\big((\hat{R}_{\mu\nu})^2\big)^2
   \,+\, d\,\sqrt{-\hat{g}}\big(\hat{R}_{\mu\nu}\big)^4
 \Big)\,.
\ee
The two Weyl-invariants on the RHS give distinct expressions
{\allowdisplaybreaks[1]
\begin{align}
\sqrt{-\hat{g}}\big((\hat{R}_{\mu\nu})^2\big)^2
&\xrightarrow{\text{EOM}}
\frac{1}{48} e^{-(d-8) \tau }  (d-2)^4 
\Big(48 (\partial \partial \tau )^4 +192 (\partial \partial \tau \partial \tau \partial \tau ) \left(\partial \partial \tau )^2
+192 (\partial \partial \tau \partial \tau \partial \tau )^2\right.\nonumber\\
	&\qquad\qquad 
	-24(d-4) (\partial \tau )^4  \left(\partial \partial \tau )^2\right.-(d-4)(d-44) (\partial \tau )^8\Big)\,,
\label{R4A}
\\
\sqrt{-\hat{g}}\big(\hat{R}_{\mu\nu}\big)^4
&\xrightarrow{\text{EOM}}\frac{1}{96} e^{-(d-8) \tau } (d-2)^4 
\Big(48 (\partial \partial \tau )^4-48 (d-12) (\partial \partial \tau \partial \tau \partial \tau ) \left(\partial \partial \tau )^2\right.\nonumber\\
	&\left.\qquad\qquad-192 (d-9) (\partial \partial \tau \partial \tau \partial \tau )^2+6 \left(d^2-22 d+96\right) (\partial \tau )^4 \left(\partial \partial \tau )^2\right.\right.\nonumber\\
	&\qquad\qquad+\left(d^3-48 d^2+626 d-2304\right) (\partial \tau )^8\Big)\,,
\label{R4B}
\end{align}
}
except in $d=8$:
\bea
\label{8dspecial}
\sqrt{-\hat{g}}\big(\hat{R}_{\mu\nu}\big)^4 
&\!\xleftrightarrow[{\tiny d=8}]{\text{~~EOM~~\,}}\!&
\frac{1}{2}\sqrt{-\hat{g}}\big((\hat{R}_{\mu\nu})^2\big)^2
\\[1mm]
 &\!\hspace{-4mm}\xrightarrow[d=8]{\text{EOM}}&
\!\hspace{-4mm}
648\Big(
(\partial \partial \tau )^4
+4 (\partial \partial \tau \partial \tau \partial \tau ) (\partial \partial \tau )^2
+ 4 (\partial \partial \tau \partial \tau \partial \tau )^2
-2 \left(\partial \partial \tau )^2\right. (\partial \tau )^4
+3 (\partial \tau )^8\Big) .
	\nonumber
\eea
So in general $d>2$, the on-shell matrix elements at $O(p^8)$ depend on two free parameters: for $d=8$ they are $\Delta a$ and the coefficient of (say) $\big(\hat{R}_{\mu\nu}\big)^4$, while for $d \ne 8$ they are the coefficients of $\big(\hat{R}_{\mu\nu}\big)^4$ and $\big((\hat{R}_{\mu\nu})^2\big)^2$. We can summarize this as
\begin{align}
S^{\partial^8} &= \int d^d x \Big[\, 
\Gamma_1 \,\big((\hat{R}_{\m\n})^2\big)^2
+\Gamma_2 \,\big(\hat{R}_{\m\n}\big)^4 
+\dots\Big]
~-~\delta_{d,8}\, \Delta a \,S_{\WZ} \, ,
\label{p8NOTgjms}
\end{align}
where the ``$\dots$" stand for terms that vanish on-shell. In 8d the amplitudes depend only on $\Delta a$ and the combination $\Gamma_{8d}\equiv 2\Gamma_{1}+\Gamma_2$.

As with  4- and 6-derivatives, the 8-derivative action can also be written in an alternative form,
\be
S^{\partial^8} ~=~ \int d^d x \bigg[\, 
\gamma\,\left(\tfrac{2}{d-8}\right)^2 e^{-(d-8)\tau/2} \square^4 e^{-(d-8)\tau/2} 
~+~ 
\tilde{\gamma}\,
\left(\tfrac{2}{d-4}\right)^2 \, e^{4\tau} \Big(  \Box^2\, e^{-\frac{d-4}{2}\, \tau}\Big)^2
+\dots\bigg]\,,~~~
\label{p8gjms}
\ee
which encodes the same $O(p^8)$ on-shell amplitudes as \reef{p8NOTgjms}, with the understanding that for $d=8$ we have $\gamma = 4\Delta a$ and  $\tilde{\gamma}=162 (\Gamma_{\text{8d}} - \frac{2}{9}\Delta a)$.  
The equality of \reef{p8NOTgjms} and \reef{p8gjms} is found by matching the coefficients of the 23 distinct terms in \reef{8derivAppBstart}-\reef{8derivAppBend} with the similar terms in \reef{p8gjms} using only 9 variables, $\gamma_1,\dots,\gamma_9$.

For $d\ne8$, the translation between coefficients in \reef{p8NOTgjms} and \reef{p8gjms} is 
\bea
  \label{Gammas}
  \Gamma_1 = \frac{36d}{(d-8)(d-2)^4} \,\gamma + \frac{4}{(d-2)^4}\, \tilde{\gamma}\,,
  ~~~~~~
  \Gamma_2 = -\frac{576}{(d-8)(d-2)^4} \,\gamma \,.
\eea
Note that the linear combination $\Gamma_{\text{8d}}$ is finite in the limit $d\to 8$.

The $n=4,5,\dots, 8$-point amplitudes at $O(p^8)$ are given in general $d$ dimensions in \reef{ADimd_p8}.
Let us here list the results for $d=8$, using $\gamma = 4 \Delta a$:
{\allowdisplaybreaks[1] \small 
\begin{align}
\mathcal{A}_{4}^{(8)}&=
  \frac{2}{81 f^{12}} \big(36 \Delta a +\tilde{\gamma }\big) 
  \big( s^4 + t^4 + u^4 \big)   
  ~=~
   \Gamma_{\text{8d}} \frac{4}{f^{12}}
  \big( s^4 + t^4 + u^4 \big)   \,,
\nonumber
\\[1mm]
\mathcal{A}_{5}^{(8)}&=
\frac{8}{243 f^{15}}
\,\Big[
\big(54\Delta a+ \tilde{\gamma } \big)\, P_{5,A}^{(8)} 
+ \tilde{\gamma } \, P_{5,B}^{(8)}
\Big]
 \,,
 \nonumber
\\[1mm]
\mathcal{A}_{6}^{(8)} &= 
\frac{8}{729 f^{18}}
\,\bigg[
  \big( 486 \Delta a+7 \tilde{\gamma } \big) \,P_{6,A}^{(8)} 
  + 2 \big(81 \Delta a+ \tilde{\gamma } \big)\,P_{6,B}^{(8)} 
  + 7 \tilde{\gamma } \, P_{6,C}^{(8)}
  + 4 \tilde{\gamma } \, P_{6,D}^{(8)} 
  \bigg]
\,,\nonumber
\\[1mm]
\mathcal{A}_{7}^{(8)}&= 
\frac{16}{243 f^{21}}
\,\bigg[  \big(324 \Delta a+7 \tilde{\gamma } \big)\,P_{7,A}^{(8)} 
  +  162 \Delta a \, P_{7,B}^{(8)} 
  + 7 \tilde{\gamma } \,P_{7,C}^{(8)} 
  + 4 \tilde{\gamma } \, P_{7,D}^{(8)} 
   \bigg]
\,,
\nonumber
\\[1mm]
\mathcal{A}_{8}^{(8)}&= 
 \frac{16}{2187 f^{24}}
\,\bigg[
  \big(14580 \Delta a+301 \tilde{\gamma } \big)\, P_{8,A}^{(8)} 
  + \big(5832 \Delta a+77 \tilde{\gamma } \big)\, P_{8,B}^{(8)} 
  +  \big(2187 \Delta a-7 \tilde{\gamma } \big)\, P_{8,C}^{(8)} 
\nonumber \\
& \hspace{2cm}
  + 189 \tilde{\gamma } \, P_{8,D}^{(8)} 
  + 126 \tilde{\gamma } \, P_{8,E}^{(8)}
   \bigg]\,,
   \label{Anp8inDim8}
\end{align}
}
where for instance,
\be
P_{5,A}^{(8)} = \sum\limits_{1\leq i<j\leq 5} s_{ij}^4\,,
\hspace{8mm}
P_{5,B}^{(8)} = s_{12}^2 s_{34}^2 + \text{perms} \,,
\ee
and the definitions of the other basis polynomials for $n\ge 6$ are given in \reef{basisP8_6pt}-\reef{basisP8_8pt}.

We noted below \reef{Swz8} that information about the anomaly cannot enter until the 5-point amplitude. This is verified by the second equality for the 4-point amplitude in \reef{Anp8inDim8} where we used 
$\tilde{\gamma}=162 (\Gamma_{\text{8d}} - \frac{2}{9}\Delta a)$ to demonstrate that the 4-dilaton amplitude indeed captures no information about the anomaly flow $\Delta{a}$.

It is a new feature in 8d, compared with 4d and 6d, that there is a non-vanishing contribution from the Weyl-invariants  at the same order in momentum as the flow in the $a$-anomaly. In 4d and 6d, the WZ action provided the unique contributions to, respectively, the $O(p^4)$ and $O(p^6)$ matrix elements. In 8d, the $O(p^8)$ dilaton matrix elements are ``polluted" by the Weyl-invariant, which does not contain information about the flow of $a$ in general. However, note that even from just the 5-dilaton amplitude one can determine $\Delta a$ and $\tilde{\gamma}$ uniquely, since there are two independent Mandelstam polynomials. The match to the higher-point amplitudes is then a strong consistency check.  We check consistency explicitly in section \ref{s:scalar8d}.

\subsection{Dilaton effective action and GJMS operators}
\label{s:gjms}
We found above that the relevant terms in the flat-space dilaton effective action were expressed in terms of $\Box^k$ up to and including $O(p^8)$. The derivation required solutions to over-constrained systems of equations. A solution could be found in each case because $\Box^k$ is the flat-space limit of the GJMS operator, $P_k$, which transforms in the following simple manner under conformal transformations:
\bea
  P_k[e^{2\sigma} g] ~=~ 
  e^{-(d/2+k) \sigma}\,P_k[g] \,e^{(d/2-k) \sigma} \,\,;
  ~~~~~~~~
  P_k[\eta]~=~\Box^k \,.
  \label{gjmsOP}
\eea
The GJMS operators are the higher-order generalizations of  
 the well-known conformal Laplacian (the Yamabe operator) 
 $P_1= \Box -\tfrac{(d-2)}{4(d-1)}R$ and the Paneitz operator $P_2 = \Box^2 +\dots$ \cite{paneitz,FradkinTseytlin}. 

Let us define
\bea 
\label{Wk}
   \mathcal{W}_k ~\equiv~\big(\tfrac{2}{d-2k}\big)^2 \,e^{-(d/2-k)\tau} P_k \,  e^{-(d/2-k)\tau} \, .
\eea
Under a Weyl-transformation, 
$\tau \to\tau +\sigma$ and $g_{\mu\nu} \to e^{2\sigma}g_{\mu\nu}$,
and it follows from \reef{gjmsOP} that 
\bea
  \mathcal{W}_k
  ~\xrightarrow{\text{Weyl}}~
  e^{-d\,\sigma} \mathcal{W}_k \,,
  ~~~~~~~(k \ne d/2)\,,
\eea
so that $\sqrt{-g}\mathcal{W}_k$ is a Weyl-invariant for $k \ne d/2$. It is the flat-space limit of $\sqrt{-g}\mathcal{W}_k$ we have encountered in the our analysis of the $O(\pa^{2k})$-derivative terms. 

The results for the $2k$-derivative actions of the previous subsections can now be summarized as
\be
   S
   = \int d^dx \,\sqrt{-g} 
   \Big[
   \tfrac{(d-2)^2}{8} f^{d-2}\,\mathcal{W}_1
   + \alpha \, \mathcal{W}_2
   + \beta \, \mathcal{W}_3
   + \gamma \,\mathcal{W}_4 
   + \tilde{\gamma} \, \tfrac{(d-4)^2}{4} \,e^{d\,\tau}\,(\mathcal{W}_2)^2
   + \dots
   \Big]\bigg|_{g_{\mu\nu}=\eta_{\mu\nu}},
   \label{Sgjms}
\ee
where the ellipses stand for 1) terms that vanish upon application on the equations of motion, and 2) terms with more than 8 derivatives. 

The normalization in \reef{Wk} was chosen such that for even $d$ we get 
\bea
  \int d^dx\, \sqrt{-g}\,\mathcal{W}_{d/2} 
  ~~~ \xrightarrow{\text{flat space}}~~~ 
  \int d^dx~\tau\, \Box^{d/2} \tau \,.
\eea
As discussed in the previous subsections, this means the coefficient of $\mathcal{W}_{d/2}$  is $(d/2) \Delta a$ for $d$ even; i.e.~$\alpha = 2 \Delta a$ for $d=4$, $\beta =  3\Delta a$ for $d=6$, and $\gamma =  4\Delta a$  for $d=8$.\footnote{One should be aware that terms like $\tau\square^{d/2}\tau$ can be produced by more than just the $\mathcal{W}_{d/2}$ operator. For instance, when $d=8$, $e^{d\tau}(\mathcal{W}_2)^2$ contains $\tau\square^4\tau$. However, the Weyl transformation of this term is compensated by the other terms produced by that operator so that the whole expression is invariant.}

Note also that
\bea
  \tfrac{(d-4)^2}{4} \,e^{d\,\tau}\,(\mathcal{W}_2)^2 
  ~=~
  \big(\tfrac{2}{d-4}\big)^2 \, e^{4\tau} \Big(  P_2\, e^{-\frac{d-4}{2}\, \tau}\Big)^2 
  ~ \xrightarrow{\text{flat space}}~
  \big(\tfrac{2}{d-4}\big)^2 \, e^{4\tau} \Big(  \Box^2\, e^{-\frac{d-4}{2}\, \tau}\Big)^2 \,.
\eea
In the limit $d \to 4$, this simply becomes $e^{4\tau}\,(\Box^2 \tau)^2$.

In this section, we have shown that in the case of flows induced by explicit breaking of the conformal symmetry, the terms that matter for extracting the on-shell dilaton amplitudes from the flat-space dilaton effective action can be written in the ``GJMS-form" \reef{Sgjms}. In the following sections, we will verify this form explicitly using the example of the RG flow of the free massive scalar field. It is tempting to propose that this is also the form of the action that matters in the conformally flat case, for example the $d$-sphere, for which the GJMS operators exist for all $k$ \cite{Juhl}. 

\setcounter{equation}{0}
\section{Example: free scalar in 8d}
\label{s:scalar8d}

The example of the free conformal scalar was studied for $d=4$ in \cite{KS} and $d=6$ in \cite{6d}.
Here we consider  $d=8$  with the purpose of testing the 8d form of the dilaton effective action derived in the previous section. We also show how the flow of the anomaly, $\Delta a$, can be separated systematically from the non-vanishing contribution of the 8-derivative Weyl-invariant $e^{8\tau}(\mathcal{W}_2)^2$ .

Consider the action for a free massive scalar in 8d, 
\begin{align}
S=\int d^8 x\left(-\frac{1}{2}(\partial\Phi)^2-\frac{1}{2}M^2\Phi^2\right) \,.
\end{align}
The presence of the mass-term operator explicitly breaks the conformal symmetry of the action. We can restore that symmetry by promoting the coupling to a scalar function of spacetime as 
\begin{align}
M^2 \rightarrow M^2 e^{-2\tau} = \lambda\Omega^{2/3} \, ,
\label{M2tau8d}
\end{align}
with $\lambda = M^2/f^2$.

Introducing a kinetic term for the compensator field $\Omega$, we write
\begin{align}
S=\int d^8 x\left(-\frac{1}{2}(\partial\Phi)^2-\frac{1}{2}(\partial\Omega)^2 -\frac{1}{2}\lambda\Omega^{2/3}\Phi^2\right) \,.
\label{8d-Sscalar}
\end{align}
The 8d scalar fields $\Phi$  and $\Omega$  have mass dimension 3, so the exponent of $2/3$ in \reef{M2tau8d} is compatible with the  coupling $\lambda$ being dimensionless.

When the compensator acquires a VEV, $\langle \Omega \rangle = f^{3}$, the mass-term for $\Phi$ is recovered. 
The fluctuation, $\varphi$, defined as $\Omega = f^{3} - \varphi$ (cf.~\reef{physdilaton}) is the physical dilaton. 
This way the explicitly broken conformal symmetry can be treated as spontaneously broken and the anomaly matching argument of KS \cite{KS,K} applies. The key difference between the truly spontaneously broken scenario and explicit breaking is that in the latter case we are free to choose the scale $f$ such that the dilaton is arbitrarily weakly coupled. 

The fractional exponent of $\Omega = f^{3} - \varphi$ means that unlike the 4d and 6d cases \cite{KS,6d}, there are an infinite number of interaction vertices $\Phi^2 \varphi^k$ in the action \reef{8d-Sscalar}:
\be
S=
\int d^8 x\left[-\frac{1}{2}(\partial\Phi)^2 
- \frac{1}{2}M^2\Phi^2 
-\frac{1}{2}(\partial\varphi)^2 
+  \frac{M^{2}}{3f}\Phi^2 \varphi 
+\frac{M^{2}}{18f^2}\Phi^2 \varphi^2 
+ \frac{2\,M^{2}}{81f^3}\Phi^2 \varphi^3+\dots\right].
\label{8d-SscalarB}
\ee
The massive $\Phi$ can be integrated out to leave the effective action for the dilaton $\varphi$. To compare with our general 8d action, an easy approach \cite{6d} is to calculate the $n$-point on-shell dilaton scattering amplitudes from \reef{8d-SscalarB} and compare with those of the general 8d dilaton effective action \reef{Anp8inDim8}. 
Taking $f \gg M$ means that the calculation is effectively 1-loop: no internal $\varphi$'s are exchanged. (This is the case of explicit breaking, and we can view the dilaton as a source \cite{6d}.) The low-energy expansion of the amplitudes in powers of external momenta results in divergent diagrams at $O(p^4)$;  the coupling of the $O(\pa^4)$ terms are renormalized in 8d, and we do not attempt to match them to the general effective action. At order $O(p^6)$ and $O(p^8)$ (and higher) the results of the 1-loop calculation are finite and a precise match is obtained up to 8-point order. 

As an example of the match, consider the 6-point $O(p^6)$ amplitudes. The calculation of the 1-loop amplitude with 6 external $\varphi$'s and an internal loop of $\Phi$'s makes use of 3-, 4-, 5-, and 6-point interactions from \reef{8d-SscalarB} and involves sums of hexagon diagrams, pentagon diagrams, 3 types of box diagrams (with topology of ``1-mass'', ``2-mass-easy" and ``2-mass-hard"), 3 types of triangle diagrams, and 2 types of bubble diagrams.\footnote{For comparison, the equivalent calculation \cite{6d} in 6d was much easier since with only a cubic vertex, the only diagram involved was the hexagon diagram.} The result can be expressed in terms of the Mandelstam basis polynomials \reef{basisPp6} as
\be
{d=8\!:}~~~~~~~~~~~~~
\mathcal{A}_6^{(6)} ~=~
\frac{2^3 M^2}{3^{10}\, 7\, (4\pi)^4 \,f^{18}}\left(16 \,P_{6,A}^{(6)} + 5 P_{6,B}^{(6)}\right)\,.~~
\label{A6p6in8d-1loop}
\ee
Comparing this with \reef{A6p6in8d} one immediately sees that the functional form matches, and we can read off $\beta= 2 M^2/(2835(4\pi)^4)$. We have explicitly checked that all the other $n=4,5,6,7,8$-point amplitudes \reef{ADimd_p4} with $d=8$ are also reproduced exactly with this value of $\beta$. While $\beta$ itself has no particular interest to us\footnote{Other than we note that $\beta$ is positive.} --- it is a model-dependent dimensionful coefficient --- the fact that we reproduce the $O(p^6)$ amplitudes is a strong consistency check on the 1-loop calculation and on the structure of the dilaton effective action at $O(\pa^6)$.

Next, move ahead to the $O(p^8)$ amplitudes which in $d=8$ contain information about the flow of the trace anomaly. Details of the calculation are given in appendix \ref{app:oneloop}; here we quote the 4- and 5-point 1-loop amplitudes:
\begin{align}
\mathcal{A}_4^{(8)}&=
\frac{17}{3\,061\,800 \, (4\pi)^4\,f^{12}}
\, \big( s^4 + t^4 + u^4 \big)\,,
\nonumber\\[2mm]
\mathcal{A}_5^{(8)}&=
\frac{16}{27 f^{15}}
\,\bigg[
  \frac{13}{777\,600\, (4\pi)^4}P_{5,A}^{(8)}+\frac{11}{5\,443\,200\, (4\pi)^4}P_{5,B}^{(8)}
\bigg]\,.
\label{Anp8inDim8-1loop}
\end{align}

Comparing $\mathcal{A}_5^{(8)}$ in \reef{Anp8inDim8-1loop} and \reef{Anp8inDim8}, we find both $\Delta a$ and $\tilde{\gamma}$ thanks to the two independent Mandelstam polynomials. The result is
\begin{align}
\label{deltaa}
\Delta a ~=~ 
\frac{23}{5\,443\,200 \,(4\pi)^4}=
\frac{23}{2^7\, 3^5\,5^2\, 7 \,(4\pi)^4}\, 
\end{align}
and 
\begin{align}
\label{tgam}
  \tilde{\gamma}~=~
   \frac{11}{151\,200 \,(4\pi)^4} 
\,.
\end{align}
This is consistent with the matching of $\mathcal{A}_4^{(8)}$ in \reef{Anp8inDim8-1loop} and \reef{Anp8inDim8} 
with 
\be
\Gamma_\text{8d}=\frac{\tilde{\gamma}}{162}+\frac{2}{9}\Delta a=
   \frac{17}{12\,247\,200 \,(4\pi)^4}\,.
\ee

Note that $\Delta a>0$ in accordance with a possible 8d $a$-theorem. Also, the coefficient of $(s^4+t^4+u^4)$ is positive as expected, $\Gamma_{8d}>0$ 
(cf.~discussion in the Introduction). As a further non-trivial consistency check, we have calculated the 1-loop $6,7,8$-point amplitudes and matched them exactly to the $O(p^8)$ amplitudes in \reef{Anp8inDim8} with the same values \reef{deltaa} and \reef{tgam}.

The UV theory is that of a free massless scalar with the corresponding Weyl anomaly $a_\UV=a_{\text{scalar,8d}}$. The mass term ignites the flow and in the deep IR the massive scalar $\Phi$ decouples. Hence the IR theory is trivial,  $a_\IR=0$. Thus we expect that $\Delta a = a_\UV - a_\IR = a_{\text{scalar,8d}}$. The anomaly $a_{\text{scalar,8d}}$ of a free conformal scalar can be calculated from the free energy on a $d$-sphere, so our value \reef{deltaa} for $\Delta a =a_{\text{scalar,8d}}$ is easily checked.  Read on.

\setcounter{equation}{0}
\section{$a_{\text{scalar,d}}$ from zeta-function regularization of the free energy}
\label{s:anomd}
The action for a free conformal scalar is 
\be
 S=\int d^8 x\, \sqrt{-g}\left(-\frac{1}{2}(\nabla\Phi)^2- \frac{d-2}{4(d-1)}R\, \Phi^2\right) \,.
\ee
Consider now the theory on a $d$-sphere $S^d$. 
In the notation of \cite{Klebanov:2011gs}, we can write the free energy 
\bea
\nonumber
   F &=& - \log |Z| 
   ~=~ \frac{1}{2} \log \det \mu_0^{-2} \Big( - \nabla^2  + \frac{d-2}{4(d-1)}R\Big) \\
   &=&
   \frac{1}{2} \sum_{n=0}^\infty m_n 
   \Big[ 
     - 2 \log (\mu_0 r_0)
    + \log(n+d/2) + \log (n-1+d/2)
   \Big]\,,
   \label{FZ}
\eea
where $r_0$ is the radius of the $S^d$, $\mu_0$ is the UV cutoff, and 
\be
m_n = \frac{(2n+d-1)(n+d-1)!}{(d-1)!\,n!}
\ee
are the multiplicities of the eigenvalues $\{\lambda_n\}_{n\ge0}$ of the conformal Laplacian on $S^d$. The coefficient of the $\log (\mu_0 r_0)$-term in \reef{FZ} is the $a$-anomaly of the free conformal scalar. 
Normalizing by the integral of the Euler density over $S^d$, we have
 \bea
   a_{\text{scalar},d} = - \frac{\sum_{n=0}^\infty m_n}{\int_{S^{d}} ~ \sqrt{g}\,E_d} \,.
   \label{a_sc_sum}
 \eea 
In our conventions \reef{eulerdens} for the Euler density this is 
\begin{align}
\int_{S^{d}} ~ \sqrt{g}\,E_d = d!~\Omega_d \,,
\label{intEd}
\end{align} 
where $\Omega_d=2\pi ^{(d+1)/2}/\Gamma\left(\frac{d+1}{2}\right)$ is the surface volume of the $d$-sphere. 

The sum in \reef{a_sc_sum} is formally divergent, but can be evaluated via zeta-function regularization. This gives $a_{\text{scalar},d\text{ odd}} = 0$ as well as the familiar values
\bea
  a_{\text{scalar,4d}} = \frac{1}{360 \, (4\pi)^2} \,,
  ~~~~~\text{and}~~~~~
  a_\text{scalar,6d} = \frac{1}{9072 \, (4\pi)^3} \,.
\eea
This method was used already in 1979 to calculate the functional determinant \reef{FZ} \cite{early}; explicit values for $d=4,6,8,10$ were given by Copeland and Toms in 1986 \cite{Copeland:1985ua}.\footnote{Table 1 in \cite{Copeland:1985ua} quotes an incorrect value for $d=12$.} More recently, Cappelli and D'Appollonio \cite{Cappelli:2000fe} extended the list of explicit values up to $d=14$. 
A compact formula for $a_{\text{scalar},d}$ was presented by Diaz \cite{Diaz:2008hy}, and it is easily translated to our conventions using \reef{intEd}:
\bea
   \text{$d$ even:}~~~~~~
   a_{\text{scalar},d} ~=~
   \frac{{\mathrm a}(d)}{d!\,\big(\frac{d}{2}\big)!\,(4\pi)^{d/2}} 
   \, ,~~~~\text{with}~~~~
    {\mathrm a}(d)=  - 
   \int_0^1 dt\, \prod_{i=0}^{d/2-1} (i^2 - t^2) \,.
   \label{anomaly-d}
\eea
For $d=4,6,\dots,20$ one finds
\bea
   {\mathrm a}(d)= \Big\{
   \tfrac{2}{15}, \tfrac{10}{21}, \tfrac{184}{45}, \tfrac{2\,104}{33}, 
   \tfrac{2\,140\,592}{1365}, 
   \tfrac{2\,512\,144}{45}, 
   \tfrac{2\,075\,529\,088}{765}, 
   \tfrac{344\,250\,108\,032}{1\,995}, 
   \tfrac{6\,884\,638\,343\,936}{495}\Big\}\,.
\eea
We have checked explicitly that these values agree with the result of zeta-function regularization of the sum \reef{a_sc_sum}. 

Note that for $d=8$, we have 
\bea 
  a_\text{scalar,8d} 
  ~=~  \frac{1}{8! \,4!\,(4\pi)^{4}}\times  \frac{184}{45} 
  ~=~ \frac{23}{5\,443\,200\, (4\pi)^4} \,.
\eea
This is in perfect agreement  with our 1-loop calculation \reef{deltaa}.\footnote{Let us note that for odd-$d$, the $O\big((r_0)^0\big)$-terms in  \reef{FZ} produce  the $F$-coefficient for a free conformal scalar; this is also evaluated using zeta-function regularization and explicit values can be found  in \cite{Klebanov:2011gs}. An approach using entanglement entropy for even-$d$ and odd-$d$ spheres was studied in  \cite{Dowker:2010bu} and \cite{Dowker:2010yj}.}

\setcounter{equation}{0}
\section{Free scalar in $d$ dimensions and the dilaton effective action}
\label{s:scalarDd}
In this section, we generalize to $d$ dimensions the example of the 8d free scalar from section \ref{s:scalar8d}. We match the dilaton effective action up to 8-derivative terms for $d=3,4,\dots,10$. Finally, we comment on the structure of higher-derivative terms.

\subsection{Free scalar in $d$ dimensions}
\label{ss:scalard}
 Consider a free massless scalar, $\Phi$. Introducing a mass term in the action,
  \begin{align}
S=\int d^d x\left(-\frac{1}{2}(\partial\Phi)^2-\frac{1}{2}M^2\Phi^2\right)\,,
\end{align}
breaks the conformal symmetry explicitly. The symmetry can be restored by promoting the mass to a spacetime dependent quantity with the introduction of a compensator field $\Omega$:
\begin{align}
S=\int d^d x\left(-\frac{1}{2}(\partial\Phi)^2 - \frac{1}{2}(\partial\Omega)^2
- \frac{1}{2}\lambda\, \Omega^{\frac{4}{d-2}}\Phi^2\right)\,.
\label{SdDim}
\end{align}
The coupling $\lambda = M^2/f^2$ is dimensionless (as is compatible with the mass-dimension $(d-2)/2$ of $d$-dimensional scalars).
To see that this makes the classical theory conformal, calculate the stress tensor from the action \reef{SdDim}, 
\begin{equation}
T_{\mu\nu}
~=~
-\frac{2}{\sqrt{-g}}\frac{\delta S}{\delta g^{\mu\nu}} 
~=~ \partial_\mu\Phi\partial_\nu\Phi + \partial_\mu\Omega\partial_\nu\Omega-\frac{1}{2}\eta_{\mu\nu}\left[(\partial\Phi)^2+(\partial\Omega)^2+\lambda\Omega^{\frac{4}{d-2}}\Phi^2\right] \, ,
\end{equation}
and improve it to
\begin{equation}
\Theta_{\mu\nu}~=~T_{\mu\nu}
~-~\frac{1}{4}\frac{d-2}{d-1}(\partial_\mu\partial_\nu-\eta_{\mu\nu}\square)(\Phi^2+\Omega^2)\,.
\end{equation}
Then upon application of the equations of motion 
\begin{equation}
\square \Phi = \lambda\,\Omega^{\frac{4}{d-2}}\,\Phi
~~\text{\quad and\quad}~~ 
\square\Omega = \lambda\, \frac{2}{d-2}\, \Omega^{\frac{4}{d-2}-1}\,\Phi^2\,
\end{equation}
one finds  $\Theta_\mu{}^{\mu} = 0$.

The model \reef{SdDim} has a moduli space along $\Omega$ when $\langle \Phi \rangle = 0$. At the origin, $\langle \Omega \rangle = 0$, the theory is conformal, but the conformal symmetry is spontaneously broken at $\langle \Omega \rangle = f^{(d-2)/2} \ne 0$. In this vacuum, the original mass term is recovered since $\lambda = M^2/f^2$, and the physical dilaton $\varphi$ is the fluctuation, $\Omega = f^{(d-2)/2} - \varphi$. 
The scale of $f$ is unrelated to $M$, so we choose $f \gg M$ to make the model perturbative, $\lambda \ll 1$. 

We are interested in calculating the $n$-point dilaton amplitudes in a leading order (in $\lambda$) low-energy expansion and comparing the results with the matrix elements extracted from the dilaton effective action in section \ref{s:action}. With $\Omega = f^{(d-2)/2} - \varphi$, the  action \reef{SdDim} gives  $n$-point interaction terms between $\Phi$ and $\varphi$ for all $n$, unless $d = 3,4,6$ when there are a finite number of terms. At leading order in $\lambda$, the dilaton scattering amplitudes are given by the 1-loop diagrams with $n$-external dilatons $\varphi$ and the massive scalar $\Phi$ running in the loop. The Feynman rule for the vertex with two $\Phi$'s and $k$ $\varphi$'s is
\begin{align}
V_k &= i(-1)^{k+1} \frac{M^2}{f^{k(d-2)/2}} \bigg(\prod\limits_{n=0}^{k-1} \Big( \frac{4}{d-2}-n\Big)\bigg) \,.
\label{vertexrule}
\end{align}
We refer the reader to appendix \ref{app:oneloop} for practical details of the 1-loop calculation. 

The results of the 1-loop calculation of the $n$-point dilaton amplitudes for the free massive scalar can be compared to the general form of the amplitudes discussed in section \ref{s:action} and listed in appendix \ref{app:ampli}. Since to $O(\pa^8)$ there are only few parameters, $\alpha$, $\beta$, $\gamma$, and $\tilde\gamma$, in the dilaton effective action \reef{Sgjms}, this provides a very non-trivial check of the structure. We find perfect consistency for $d=3,4,5,\dots,10$. Specifically:

\begin{itemize}
\item At $O(p^4)$ and $O(p^6)$ we have checked the form of the action \reef{Sgjms} for $d=3,4, \dots ,7$ (and also $d=8,9$ for $O(p^6)$) by matching the amplitudes with $n=4,5,6,7,8$ external dilatons.  To illustrate the non-triviality of the match, note that for $O(p^6)$ this requires matching the coefficients of a total of 8 independent momentum polynomials of \reef{ADimd_p6} in terms of just a single free parameter, $\beta$.
\item At $O(p^8)$, we have matched the $d=3,4, \dots ,10$ dilaton $n$-point amplitudes with $n=4,5,6,7$. This requires matching the coefficients of 11 independent Mandelstam polynomials in \reef{ADimd_p8} using just two parameters $\gamma$ and $\tilde{\gamma}$.\footnote{Note that the constrained 3d kinematics leave fewer independent Mandelstam polynomials than for $d>3$.}  For $d=8$, we also matched the 8-point amplitude with its 5 independent momentum polynomials. And as noted in sections \ref{s:scalar8d}-\ref{s:anomd}, the 8d anomaly flow $\Delta{a}=\gamma/4$ for the free massive scalar was correctly reproduced by this calculation.
\end{itemize}
Thus the 1-loop calculation for the free massive scalar offers a highly non-trivial check of the dilaton effective action.  

In Table \ref{table1}, we summarize the results for the coefficients $\alpha$, $\beta$, $\gamma$, and $\tilde\gamma$ in $d=3,4,5,\dots,10$. Note that the boxed values in the table correspond to the anomaly flows
\bea
  \nonumber
  \Delta{a}_\text{4d} &=& \frac{1}{2}\alpha_\circ (4 \pi)^{-2} 
  ~=~ \frac{1}{360 (4 \pi)^2}\,,\\
  \nonumber
  \Delta{a}_\text{6d} &=& \frac{1}{3}\beta_\circ (4 \pi)^{-3} 
  ~=~ \frac{1}{9072 (4 \pi)^3}\,,\\
  \Delta{a}_\text{8d} &=& \frac{1}{4}\gamma_\circ (4 \pi)^{-4} 
  ~=~ \frac{23}{5\,443\,200 (4 \pi)^4}\,.
\label{anomtable}
\eea

\begin{table}[t]
\begin{displaymath}
\begin{array}{|c|c|c|c|c|}
\hline
d & \alpha_\circ  & \beta_\circ & \gamma_\circ & \tilde\gamma_\circ \\[1pt]
\hline
3 & \phantom{\Big(} \!\!\!\!\! \frac{1}{960 \, M} 
& -\frac{1}{43\,008 \,M^3}
& -\frac{1}{92\,160\, M^5}
& \frac{1}{7\,680 \,M^5}
\\[1mm]
4 &
\boxed{\scriptstyle \frac{1}{180} }
& \frac{1}{7\,560 \,M^2}
& -\frac{1}{85\,050 \, M^4} 
& \frac{11}{37\,800 \,M^4}
\\[1mm]
5 & \frac{M}{480}
& \frac{67}{967\,680\, M}
& \frac{1}{1\,935\,360 \, M^3}
& \frac{1}{18\,432 \,M^3}
 \\[1mm]
6 & \frac{\,M^2}{90} 
& \boxed{\scriptstyle \frac{1}{3\,024}}
& \frac{1}{113\,400 \,M^2}
& \frac{1}{7\, 560 \,M^2}
\\[1mm]
7 & \frac{\,M^3}{144} 
& \frac{61\, M}{483\,840} 
& \frac{1}{272\,160\, M}
& \frac{13}{483\,840\, M}
\\[1mm]
8 &\text{\small div} 
& \frac{2\,M^2}{2\,835}
& \boxed{\scriptstyle \frac{23}{1\,360\,800}}
& \frac{11}{151\,200}
\\[1mm]
9 & 
& \frac{113\,M^3}{241\,920}
& \frac{47 \,M}{7\,257\,600}
& \frac{41 \,M}{2\,419\,200}
\\[1mm]
10 
& 
& \text{\small div} 
& \frac{151 \,M^2}{4\,082\,400}
& \frac{\,M^2}{18\,144}
\\[1mm] \hline
\end{array}
\end{displaymath}
\caption{Results for the coefficients $\alpha$, $\beta$, $\gamma$, and $\tilde\gamma$ of the effective action \reef{Sgjms} for the case of the $d$-dimensional free massive scalar flow. The subscript ${}_\circ$ in the table indicates that a factor of $(4 \pi)^{-\lfloor d/2\rfloor}$ was taken out, e.g.~$\alpha = \alpha_\circ (4 \pi)^{-\lfloor d/2\rfloor}$. The label ``div'' indicates that the 1-loop scalar integral diverges at and beyond this order. The boxed results are those encoding the $d=4,6,8$ anomaly flows for the free massive scalar; see \reef{anomtable}. Terms with negative mass-dimension are not needed for our study of RG flows, but we include them here to illustrate that the amplitudes match even in those higher-derivative cases}
\label{table1}
\end{table}

The successful match of the amplitudes for the free massive scalar and the simplicity of the dilaton effective action $S$ in the form \reef{Sgjms} encourages us to speculate about the higher-derivative terms in $S$. We outline some ideas and tests of this in the following section.

\subsection{Higher-order effective action?}
\label{s:higher}
In section \ref{s:gjms} we wrote the flat space dilaton effective action 
\be
   S
   = \int d^dx \,\sqrt{-g} 
   \Big[
   \tfrac{(d-2)^2}{8} f^{d-2}\,\mathcal{W}_1
   + \alpha \, \mathcal{W}_2
   + \beta \, \mathcal{W}_3
   + \gamma \,\mathcal{W}_4 
   + \tilde{\gamma} \, \tfrac{(d-4)^2}{4} \,e^{d\,\tau}\,(\mathcal{W}_2)^2
   + \dots
   \Big]\bigg|_{g_{\mu\nu}=\eta_{\mu\nu}},
   \label{Sgjms2}
\ee
with the ellipses standing for terms that vanish upon application on the equations of motion, plus terms with more than 8 derivatives. Recall that its definition in terms of the GJMS operators $P_k$ in \reef{gjmsOP} makes the behavior of $\mathcal{W}_k\equiv\big(\tfrac{2}{d-2k}\big)^2 \,e^{-(d/2-k)\tau} P_k \,  e^{-(d/2-k)\tau}$ under Weyl transformations particularly simple, $\mathcal{W}_k \xrightarrow{\text{Weyl}} e^{-d\,\sigma} \mathcal{W}_k$. This ensures Weyl-invariance of the action \reef{Sgjms2}, except for $d=2k$ where it produces the correct trace anomaly; the relation between the coefficients in \reef{Sgjms2} and the anomaly flow was given in \reef{anomtable}.

The simplicity of \reef{Sgjms2} encourages a guess for the 10-derivative terms, namely
\be
S^{\pa^{10}}
   = \int d^dx \,\sqrt{-g} 
   \Big[
    \delta \, \,\mathcal{W}_5 
   + \tilde{\delta} \, \tfrac{(d-4)(d-6)}{4} \,e^{d\,\tau}\, \mathcal{W}_2\, \mathcal{W}_3
   + \dots
   \Big]\bigg|_{g_{\mu\nu}=\eta_{\mu\nu}}\!\!.
   \label{S10gjms}
\ee
The ``\dots" denote terms that vanish on-shell.\footnote{Note that terms with $\mathcal{W}_1$ vanish on-shell.}
 For $d=3,4,\dots,10$, we have checked explicitly that the $4,5,6$-point amplitudes produced by the action \reef{S10gjms} are matched exactly by the $O(p^{10})$ dilaton amplitudes produced by the free massive scalar 1-loop computation. For each $d$, this requires the coefficient of 9 distinct Mandelstam polynomials to be matched using just two constants, $\delta$ and $\tilde{\delta}$, and it is therefore encouraging that this guess works. However, those two constants are not sufficient to match the 7-point amplitude; the guess in \reef{S10gjms} is incomplete.  Moreover, following the pattern of the lower-order terms, we would expect the anomaly flow in $d=10$ to be encoded as $\Delta{a} = \frac{1}{5} \delta$. Instead, we find that $\frac{1}{5} \delta \neq \Delta a_\text{10d}$. Hence at least one other term is required in \reef{S10gjms}.

A further complication (or feature) arises for the 12-derivative terms. The GJMS construction suggests that we can write 
\be
S^{\pa^{12}}\!\!
   =\!\! \int d^dx \,\sqrt{-g} 
   \Big[
    \epsilon_1 \,\mathcal{W}_6 
   + \epsilon_2  \tfrac{(d-6)^2}{4} \, e^{d\,\tau}\,(\mathcal{W}_3)^2
   + \epsilon_3  \tfrac{(d-4)(d-8)}{4} \, e^{d\,\tau}\,\mathcal{W}_2 \mathcal{W}_4
   + \epsilon_4  \tfrac{(d-4)^3}{8} \, e^{2d\,\tau}\, (\mathcal{W}_2)^3
   + \dots
   \Big]\bigg|_{g_{\mu\nu}=\eta_{\mu\nu}}
   \label{S12gjms}
\ee
However, this \emph{cannot} be the full answer, because starting at $O(p^{12})$, the 4-point amplitude has two independent Mandelstam polynomials.\footnote{This follows the same structure as the matrix elements of the candidate counterterm operators $D^{2k}R^4$ in supergravity; see for example Table 1 in 
\cite{Elvang:2010jv}.} For example in $d=12$ we find for the free massive scalar flow:
\be
  \mathcal{A}_4^{(12)} 
  ~=~ 
  \frac{1}{2^5\,3^3\,5^7\,7^2\, 11^1 \,13^1} 
  \Big( 
     -7 s^2 t^2 u^2 + 2250 (s^6 + t^6 + u^6)
  \Big) \,.
\ee
The polynomial $s^6 + t^6 + u^6$ can be produced by the terms in \reef{S12gjms}, but $s^2 t^2 u^2$ cannot. This means that new structures appear in the effective action at 12-derivative order. This may be evidence for the existence of a new class of curved-space GJMS-type operators whose ``leading" components are not $\Box^k$ but are perhaps composed of various contractions of $G_{\mu\nu\rho}=(\nabla_\mu \nabla_\nu \nabla_\rho)$.
For instance, a term in the action that also produces $s^2t^2u^2$ could be:
\be
\varphi ~ G_{\mu\nu\rho}~ \varphi ~ G^{\mu\nu\rho} ~ G^{\sigma \lambda \kappa} ~ \varphi ~ G_{\sigma \lambda \kappa}~ \varphi\,.
\ee
Such new operators may also enter the 10-derivative action and account for the mismatch of the 1-loop amplitudes predicted by extrapolating the GJMS-type action. It would be interesting to explore further the connections between RG flows, conformal geometry, functional determinants, and the $a$- and $F$-theorems.


\setcounter{equation}{0}
\section*{Acknowledgements}
We are grateful to Ratin Akhoury, Dan Freedman, Janet Hung, Michael Kiermaier, Zohar Komargodski, Rob Myers, Stephan Stieberger, Stefan Theisen, and Arkady Tseytlin for useful discussions. 

HE is supported by NSF CAREER Grant PHY-0953232, and in part by the US Department of Energy under DoE Grant 
\#DE-SC0007859. TMO is supported by a Regents Graduate Fellowship at the University of Michigan.

\appendix

\setcounter{equation}{0}
\section{Euler density and the WZ action}
\label{app:euler8d}

We begin by constructing the $d=2k$ dimensional Euler density for a metric $g_{\mu\nu}$ from its definition:
\begin{align}
E_{2k}(g_{\mu\nu}) &= 
\frac{1}{2^k}\,
R_{\mu_1\nu_1}{}^{\rho_1\sigma_1} \dots R_{\mu_k\nu_k}{}^{\rho_k\sigma_k}  
\,\epsilon_{\rho_1\sigma_1\dots\rho_k\sigma_k} 
\,\epsilon^{\mu_1\nu_1\dots\mu_k\nu_k}\nonumber\\
&= \frac{d!}{2^k}
R_{\mu_1\nu_1}{}^{\rho_1\sigma_1} \dots R_{\mu_k\nu_k}{}^{\rho_k\sigma_k}  
\delta^{\mu_1}_{[\rho_1}\dots\delta^{\nu_k}_{\sigma_k]}\,.
\end{align}
The Euler density can be written in terms of Ricci scalars and Ricci tensors plus terms involving the Weyl tensor using the identity \be
\label{RiemannWeyl}
R_{\mu\nu\rho\sigma} = W_{\mu\nu\rho\sigma}+\frac{2}{d-2}\left(g_{\mu[\rho}R_{\sigma]\nu}-g_{\nu[\rho}R_{\sigma]\mu}\right)-\frac{2}{(d-2)(d-2)}Rg_{\mu[\rho}g_{\sigma]\nu} \,,
\ee
where $W_{\mu\nu\rho\sigma}$ is the Weyl tensor. The 8-dimensional Euler density for conformally flat space is then:
\be
E_{8}(g_{\mu\nu})
= -\frac{16}{9} \big(R_{\mu\nu}\big)^4 +\frac{8}{9} \big((R_{\mu\nu})^2\big)^2 +\frac{32}{21}R\big(R_{\mu\nu}\big)^3 -\frac{344}{441}
R^2\big(R_{\mu\nu}\big)^2+\frac{208}{3087}R^4 + \text{Weyl-terms}\,.~
\ee
We are interested in the Wess-Zumino action in a flat background, so we pick $e^{-2t\tau}\eta_{\mu\nu}$ and integrate $t$ over the interval $[0,1]$:
\bea
S_\WZ&=& 
\int d^8 x \int_0^1 dt\, \tau\, E_8(e^{-2t\tau}\eta_{\mu\nu})\nonumber\\
&=&
48 \int d^8  x  \Big[
  3 (\square ^2\tau ) (\partial \tau )^4
  +6 (\square \tau )^3 (\partial \tau )^2
  +36 (\square \tau )^2 (\partial \partial \tau \partial \tau \partial \tau )
  +16 (\square \tau ) (\partial \partial \partial \tau \partial \tau \partial \tau \partial \tau )
\nonumber\\
&& \hspace{2cm}
	-12 (\square \tau ) (\partial \partial \tau )^2 (\partial \tau )^2
	-24 (\partial \partial \tau \partial \tau \partial \tau ) (\partial \partial \tau )^2
	\nonumber\\
	&& \hspace{2cm}
	+12 (\square \tau )^2 (\partial \tau )^4
	-12 (\partial \partial \tau )^2 (\partial \tau )^4
	-20 (\square \tau ) (\partial \tau )^6
	+15 (\partial \tau )^8 \Big]\,.
\eea

\setcounter{equation}{0}
\section{Diff$\times$Weyl invariants in $d$ dimensions}
\label{app:WeylInvariants}
The diff$\times$Weyl invariants in flat space are constructed as curvature scalars of the `hatted' metric $\hat{g}_{\mu\nu} = e^{-2\tau} \,\eta_{\mu\nu}$. We need only work with the scalars constructed from the Ricci tensor, Ricci scalar and covariant derivatives thereof, since the Riemann tensor can be eliminated with \reef{RiemannWeyl}.

The results for the diff$\times$Weyl invariants are expressed in terms of the dilaton $\tau$ and its derivatives. Since these terms will appear in the dilaton effective action, we use partial integration to simplify the expressions. This is indicated with ``$\xrightarrow{\text{PI}}$" below. The 2-derivative terms were discussed in section \ref{s:action}; here we present the details for 4, 6, and 8-derivative Weyl invariants.

\subsection*{4 derivatives}
There are two 4-derivative invariants:
\begin{align}
\sqrt{-\hat{g}}\hat{R}^2
&~\xrightarrow{\text{PI}}~
e^{-(d-4) \tau } \Big(4 (d-1)^2 (\square \tau )^2-4 (d-1)^2 (d-2) (\square \tau ) (\partial \tau )^2 
\nonumber\\
	&\hspace{2.6cm}
	+ (d-1)^2(d-2)^2 (\partial \tau )^4\Big)\,,
\\[1mm]
\sqrt{-\hat{g}}\,\big(\hat{R}_{\mu\nu}\big)^2
&~\xrightarrow{\text{PI}}~
e^{-(d-4) \tau } \Big(d\,(d-1) (\square \tau )^2-\frac{1}{2} (d-2) (3 d^2-8 d+8) (\square \tau ) (\partial \tau )^2
\nonumber\\
	&\hspace{2.6cm}
+\frac{1}{2} (d-2)^2 (d^2-4 d+6) (\partial \tau )^4\Big)\,.
\label{4derivAppB}
\end{align}

\subsection*{6 derivatives}
In general dimension $d$, there are 4 independent 6-derivative invariants:
{\allowdisplaybreaks[4] \small
\begin{align}
\label{6derivAppBa}
\sqrt{-\hat{g}}\hat{R}^3 &\xrightarrow{\text{PI}} 
e^{-(d-6) \tau } (d-1)^3 \Big(8 (\square \tau )^3
-12 (d-2) (\square \tau )^2 (\partial \tau )^2
+6 (d-2)^2 (\square \tau ) (\partial \tau )^4
-(d-2)^3 (\partial \tau )^6 \Big)\,,
	\\
	\label{6derivAppBb}
\sqrt{-\hat{g}}\hat{R}\big(\hat{R}_{\mu\nu}\big)^2 &\xrightarrow{\text{PI}} 
\frac{1}{2} e^{-(d-6) \tau } (d-1) 
\Big(4 (3d-4) (\square \tau )^3
+4(d-2)^2 (\square \tau ) (\partial \partial \tau )^2
+8(d-2)^2 (\square \tau )(\partial \partial \tau \partial \tau \partial \tau )
  \nonumber\\
	&  \hspace{1cm}
	-2 (d-2) (11 d-16) (\square \tau )^2 (\partial \tau )^2
	-2(d-2)^3 (\partial \partial \tau )^2 (\partial \tau )^2
	 \nonumber\\
	& \hspace{1cm}
	+(13 d-18)(d-2)^2 (\square \tau ) (\partial \tau )^4
	-(d-2)^3 (3 d-8) (\partial \tau )^6\Big)\,,
	\\
	\label{6derivAppBc}
\sqrt{-\hat{g}}\hat{R}\hat{\square}\hat{R} &\xrightarrow{\text{PI}}
\frac{1}{2} e^{-(d-6) \tau } (d-1)^2 
\Big(8 (\square ^2\tau ) (\square \tau )
+12 (d-2) (\square \tau )^3
-16(d-2) (\square \tau ) (\partial \partial \tau )^2
\nonumber\\
	& \hspace{1cm}
	+8 (d-10)(d-2) (\square \tau ) (\partial \partial \tau \partial \tau \partial \tau )
	-16 (d^2-6d-10) (\square \tau )^2 (\partial \tau )^2 
	\nonumber\\
	& \hspace{1cm}
	+4(d-2)^2 (\partial \partial \tau )^2 (\partial \tau )^2
	+(5d^2-20d-12)(d-2) (\square \tau ) (\partial \tau )^4
	\nonumber\\
	& \hspace{1cm}-(d^2-8d+20)(d-2)^2 (\partial \tau )^6 \Big)\,,
	\\
	\label{6derivAppBd}
\sqrt{-\hat{g}}\big(\hat{R}_{\mu\nu}\big)^3 &\xrightarrow{\text{PI}}
\frac{1}{8} e^{-(d-6) \tau } \Big(-4 (d^3-6 d^2+4 d+4) (\square \tau )^3
+12\,d\,(d-2)^2 (\square \tau )(\partial \partial \tau )^2 \nonumber\\
	& 
	 \hspace{1cm}
	+24\, d\,(d-2)^2 (\square \tau ) (\partial \partial \tau \partial \tau \partial \tau ) 
	+4 (d-2) (2 d^3-17 d^2+26 d-6) (\square \tau )^2 (\partial \tau )^2  
	\nonumber\\
	& \hspace{1cm}
	-4 (2d-3)(d-2)^3 (\partial \partial \tau )^2 (\partial \tau )^2
	 -(5 d^3-55 d^2+126 d-96)(d-2)^2 (\square \tau ) (\partial \tau )^4 
	 \nonumber\\
	& \hspace{1cm}
	+(d-2)^4 (d^2-13d+32) (\partial \tau )^6\Big)\,.
\end{align}
}%

\subsection*{8 derivatives}
At the level of 8 derivatives, we have found 9 independent diff$\times$Weyl invariants:
\bea
\label{listR8der}
R^4,~\,
R^2(R_{\mu\nu})^2,~\,
R({R}_{\mu\nu})^3,~\,
\big(({R}_{\mu\nu})^2\big)^2,~\,
({R}_{\mu\nu})^4,~\,
({\square}{R})^2,~\,
({\square}{R}_{\mu\nu})^2,~\,
{R}({\nabla}_\mu{R})^2,  ~\,
({R}_{\mu\nu})^2{\square}{R},\,~~
\eea
where $\sqrt{-\hat{g}}$ is implicit and we use the shorthand notation
\bea
\nonumber \label{contractions}
  &&(R_{\mu\nu})^2 \equiv R_{\mu\nu}R^{\mu\nu}\,,~~~
  ({R}_{\mu\nu})^3 \equiv R^\mu{}_\nu R^\nu{}_\rho R^\rho{}_\mu\,,~~~
  ({R}_{\mu\nu})^4 \equiv R^\mu{}_\nu R^\nu{}_\rho R^\rho{}_\sigma R^\sigma{}_\mu\,,~~~\\[1mm]
  &&({\nabla}_\mu{R})^2 \equiv ({\nabla}_\mu{R})({\nabla}^\mu{R})\,,~~~
  ({\square}{R}_{\mu\nu})^2 \equiv (\Box  R_{\mu\nu}) (\Box R^{\mu\nu})\,.
\eea

\vspace{2mm}
Due to the complexity of the off-shell expressions in general $d$ dimensions, we have opted to display only the $d=8$ forms:
{\allowdisplaybreaks[2] \footnotesize
\begin{align}
\hat{R}^4&\xrightarrow{\text{PI}}
38416 \Big[ 
(\square \tau )^4
-12 (\square \tau )^3 (\partial \tau )^2
+54 (\square \tau )^2 (\partial \tau )^4
-108 (\square \tau ) (\partial \tau )^6
+81 (\partial \tau )^8 \Big] \label{8derivAppBstart}
\,, \\[2pt]
\hat{R}^2\big(\hat{R}_{\mu\nu}\big)^2&\xrightarrow{\text{PI}}
784 \Big[
5 (\square \tau )^4
+9 (\square \tau )^2 (\partial \partial \tau )^2
+18 (\square \tau )^2 (\partial \partial \tau \partial \tau \partial \tau )
-69 (\square \tau )^3 (\partial \tau )^2 
-54 (\square \tau ) (\partial \partial \tau )^2 (\partial \tau )^2
\nonumber \\
	&\quad
+342 (\square \tau )^2 (\partial \tau )^4
+81 (\partial \partial \tau )^2 (\partial \tau )^4
-108 (\square \tau ) (\partial \partial \tau \partial \tau \partial \tau ) (\partial \tau )^2
-756 (\square \tau ) (\partial \tau )^6
+567 (\partial \tau )^8
\Big] \,,
\\[2pt]
\hat{R}\big(\hat{R}_{\mu\nu}\big)^3&\xrightarrow{\text{PI}}
28 \Big[
13 (\square \tau )^4
\!+\!108 (\square \tau ) (\partial \partial \tau \partial \partial \tau \partial \partial \tau )
\!+\!54 (\square \tau )^2 (\partial \partial \tau )^2
\!+\!648 (\square \tau ) (\partial \partial \partial \tau \partial \tau \partial \tau \partial \tau ) 
\!+\!1728 (\square \tau )^2 (\partial \partial \tau \partial \tau \partial \tau )
\nonumber \\
	&\quad
-972 (\partial \partial \tau \partial \tau \partial \tau ) (\partial \partial \tau )^2
+162 (\square ^2\tau) (\partial \tau )^4
+276 (\square \tau )^3 (\partial \tau )^2 
-1620 (\square \tau ) (\partial \partial \tau )^2 (\partial \tau )^2
+972 (\square \tau )^2 (\partial \tau )^4
\nonumber \\
	&\quad
+1215 (\partial \partial \tau )^2 (\partial \tau )^4 
-1620 (\square \tau ) (\partial \partial \tau \partial \tau \partial \tau ) \left(\partial \tau )^2\right.
-3024 (\square \tau ) (\partial \tau )^6
+2268 (\partial \tau )^8
\Big]\,,
\\[2pt]
\big((\hat{R}_{\mu\nu})^2\big)^2&\xrightarrow{\text{PI}}
16 \Big[
25 (\square \tau )^4
+90 (\square \tau )^2 (\partial \partial \tau )^2
+81 (\partial \partial \tau )^4
+180 (\square \tau )^2 (\partial \partial \tau \partial \tau \partial \tau )
+324 (\partial \partial \tau \partial \tau \partial \tau ) (\partial \partial \tau )^2
 \nonumber \\
	&\quad
-390 (\square \tau )^3 (\partial \tau )^2
-702 (\square \tau ) (\partial \partial \tau )^2 (\partial \tau )^2
+324 (\partial \partial \tau \partial \tau \partial \tau )^2
+2151 (\square \tau )^2 (\partial \tau )^4
 \nonumber \\
	&\quad
+1134 (\partial \partial \tau )^2 (\partial \tau )^4 
-1404 (\square \tau ) (\partial \partial \tau \partial \tau \partial \tau ) (\partial \tau )^2
-5292 (\square \tau ) (\partial \tau )^6
+3969 (\partial \tau )^8
\Big]\,,
\\[2pt]
\big(\hat{R}_{\mu\nu}\big)^4&\xrightarrow{\text{PI}}
8 \Big[
31 (\square \tau )^4
+324 (\square \tau ) (\partial \partial \tau \partial \partial \tau \partial \partial \tau )+81 (\partial \partial \tau )^4
-135 (\square \tau )^2 (\partial \partial \tau )^2
+1944 (\square \tau ) (\partial \partial \partial \tau \partial \tau \partial \tau \partial \tau )
 \nonumber \\
	&\quad
+4590 (\square \tau )^2 (\partial \partial \tau \partial \tau \partial \tau )
-2592 (\partial \partial \tau \partial \tau \partial \tau ) (\partial \partial \tau )^2
+486 (\square ^2\tau) (\partial \tau )^4
+1221 (\square \tau )^3 (\partial \tau )^2
 \nonumber \\
	&\quad
-3240 (\square \tau ) (\partial \partial \tau )^2 (\partial \tau )^2
+324 (\partial \partial \tau \partial \tau \partial \tau )^2
+189 (\square \tau )^2 (\partial \tau )^4
+1296 (\partial \partial \tau )^2 (\partial \tau )^4
 \nonumber \\
	&\quad
-1620 (\square \tau ) (\partial \partial \tau \partial \tau \partial \tau ) (\partial \tau )^2
-1512 (\square \tau ) (\partial \tau )^6
+1134 (\partial \tau )^8
\Big]\,,
\\[2pt]
\big(\hat{\square}\hat{R}\big)^2&\xrightarrow{\text{PI}}
\frac{196}{3} \Big[
3 (\square ^2\tau )^2
+48 (\square ^2\tau ) (\square \tau )^2
+48 (\square \tau ) (\partial \partial \partial \tau )^2
-60 (\square ^2\tau) (\partial \partial \tau )^2
\!+\!140 (\square \tau )^4
\!+\!192 (\square \tau ) (\partial \partial \partial \tau \partial \partial \tau \partial \tau ) \nonumber \\
	&\quad
+384 (\square \tau ) (\partial \partial \tau \partial \partial \tau \partial \partial \tau )
-120 (\square ^2\tau) (\partial \partial \tau \partial \tau \partial \tau ) 
+108 (\partial \partial \tau )^4
-456 (\square \tau )^2 (\partial \partial \tau )^2
-84 (\square ^2\tau ) (\square \tau ) (\partial \tau )^2
 \nonumber \\
	&\quad
+576 (\square \tau ) (\partial \partial \partial \tau \partial \tau \partial \tau \partial \tau) +624 (\square \tau )^2 (\partial \partial \tau \partial \tau \partial \tau )
-432 (\partial \partial \tau \partial \tau \partial \tau ) (\partial \partial \tau )^2
+288 (\square ^2\tau ) (\partial \tau )^4
\tau ) 
 \nonumber \\
	&\quad
-216 (\square \tau )^3 (\partial \tau )^2 
-72 (\square \tau ) (\partial \partial \tau )^2 (\partial \tau )^2
+432 (\partial \partial \tau \partial \tau \partial \tau )^2
+2028 (\square \tau )^2 (\partial \tau )^4
-864 (\partial \partial \tau )^2 (\partial \tau )^4
 \nonumber \\
	&\quad
+3600 (\square \tau ) (\partial \partial \tau \partial \tau \partial \tau ) (\partial \tau )^2
-2304 (\square \tau ) (\partial \tau )^6
+1728 (\partial \tau )^8
\Big]\,,
\\[2pt]
\big(\hat{\square}\hat{R}_{\mu\nu}\big)^2&\xrightarrow{\text{PI}}
\frac{2}{3} \Big[
84 (\square ^2\tau )^2
+777 (\square ^2\tau ) (\square \tau )^2
+156 (\square \tau ) (\partial \partial \partial \tau )^2
-762 (\square ^2\tau ) (\partial \partial \tau )^2
+662 (\square \tau )^4
 \nonumber \\
	&\quad
-8448 (\square \tau ) (\partial \partial \partial \tau \partial \partial \tau \partial \tau )
-1272 (\square \tau ) (\partial \partial \tau \partial \partial \tau \partial \partial \tau )
-1524 (\square ^2\tau ) (\partial \partial \tau \partial \tau \partial \tau )
+2376 (\partial \partial \tau )^4
 \nonumber \\
	&\quad
-2454 (\square \tau )^2 (\partial \partial \tau )^2
-3432 (\square ^2\tau ) (\square \tau ) (\partial \tau )^2
-8064 (\square \tau ) (\partial \partial \partial \tau \partial \tau \partial \tau \partial \tau )
-25\,080 (\square \tau )^2 (\partial \partial \tau \partial \tau \partial \tau )
 \nonumber \\
	&\quad
-5616 (\partial \partial \tau \partial \tau \partial \tau ) (\partial \partial \tau )^2
+2016 (\square ^2\tau ) (\partial \tau )^4
-15012 (\square \tau )^3 (\partial \tau )^2
+7488 (\square \tau ) (\partial \partial \tau )^2 (\partial \tau )^2
 \nonumber \\
	&\quad
-17712 (\partial \partial \tau \partial \tau \partial \tau )^2
+45444 (\square \tau )^2 (\partial \tau )^4
-540 (\partial \partial \tau )^2 (\partial \tau )^4 \nonumber \\
	&\quad
+59328 (\square \tau ) (\partial \partial \tau \partial \tau \partial \tau ) (\partial \tau )^2
-64512 (\square \tau ) (\partial \tau )^6
+48384 (\partial \tau )^8
\Big]\,,
\\[2pt]
\hat{R}\big(\hat{\nabla}_\mu\hat{R}\big)^2&\xrightarrow{\text{PI}}
-\frac{1372}{3} \Big[
3 (\square ^2\tau ) (\square \tau )^2
+14 (\square \tau )^4
-18 (\square \tau )^2 (\partial \partial \tau )^2
-18 (\square ^2\tau ) (\square \tau ) (\partial \tau )^2
-108 (\square \tau )^2 (\partial \partial \tau \partial \tau \partial \tau )
 \nonumber \\
	&\quad
+27 (\square ^2\tau ) (\partial \tau )^4
-150 (\square \tau )^3 (\partial \tau )^2
+108 (\square \tau ) (\partial \partial \tau )^2 (\partial \tau )^2
+594 (\square \tau )^2 (\partial \tau )^4
 \nonumber \\
	&\quad
-162 (\partial \partial \tau )^2 (\partial \tau )^4
+648 (\square \tau ) (\partial \partial \tau \partial \tau \partial \tau ) (\partial \tau )^2
-864 (\square \tau ) (\partial \tau )^6
+648 (\partial \tau )^8
\Big]\,,
\\
\big(\hat{R}_{\mu\nu}\big)^2\hat{\square}\hat{R}&\xrightarrow{\text{PI}}
\frac{56}{3} \Big[
15 (\square ^2\tau ) (\square \tau )^2
+27 (\square ^2\tau ) (\partial \partial \tau )^2
+70 (\square \tau )^4
+432 (\square \tau ) (\partial \partial \partial \tau \partial \partial \tau \partial \tau )   
+54 (\square ^2\tau ) (\partial \partial \tau \partial \tau \partial \tau )
 \nonumber \\
	&\quad
-162 (\partial \partial \tau )^4
+180 (\square \tau )^2 (\partial \partial \tau )^2
-117 (\square ^2\tau ) (\square \tau ) (\partial \tau )^2
+432 (\square \tau ) (\partial \partial \partial \tau \partial \tau \partial \tau \partial \tau)
+216 (\square \tau )^2 (\partial \partial \tau \partial \tau \partial \tau ) 
\nonumber \\
	&\quad
+297 (\square ^2\tau ) (\partial \tau )^4
-696 (\square \tau )^3 (\partial \tau )^2
-108 (\square \tau ) (\partial \partial \tau )^2 (\partial \tau )^2
+648 (\partial \partial \tau \partial \tau \partial \tau )^2
+3888 (\square \tau )^2 (\partial \tau )^4
 \nonumber \\
	&\quad
-486 (\partial \partial \tau )^2 (\partial \tau )^4
+3888 (\square \tau ) (\partial \partial \tau \partial \tau \partial \tau ) (\partial \tau )^2
-6048 (\square \tau ) (\partial \tau )^6
+4536 (\partial \tau )^8 
\Big]\,.\label{8derivAppBend}
\end{align}
}%

\setcounter{equation}{0}
\section{Dilaton amplitudes in $d$-dimensions }
\label{app:ampli}
Here we list the dilaton amplitudes at $O(p^4)$, $O(p^6)$, and $O(p^8)$ for $n=4,5,\dots,8$ as derived from the general $d$-dimensional dilaton effective action \reef{Sgjms}:

\vspace{2mm}
\noindent $\blacktriangleright$  $O(p^4)$ amplitudes:
\bea
\mathcal{A}_4^{(4)}&=& \frac{32}{(d-2)^4} 
\frac{\alpha}{f^{2d-4}}\,\big( s^2 + t^2 + u^2\big)
\,,
\hspace{14mm}
\mathcal{A}_5^{(4)}~=~\frac{32 \,d\, }{(d-2)^5} 
\frac{\alpha}{f^{5 d/2-5}}\,P_5^{(4)} 
\,,\nonumber\\[1pt]
\mathcal{A}_6^{(4)}&=& \frac{32 \,d\, (3d-2)}{(d-2)^6}
\frac{\alpha}{f^{3d-6}}\,P_6^{(4)}  \,,
\hspace{23mm}
\nonumber
\mathcal{A}_{7}^{(4)}~=~ \frac{128 \,d\,(d-1) (3d-2)}{(d-2)^7}
\frac{\alpha}{f^{7d/2-7}}\,P_{7}^{(4)} 
\,,\nonumber\\[1pt]
\mathcal{A}_{8}^{(4)}&=& 
\frac{128 \,d\,(d-1) (3d-2)(5d-6)}{(d-2)^8}
\frac{\alpha}{f^{4d-8}}\,P_{8}^{(4)} \,,
\label{ADimd_p4}
\eea
where $P_{n}^{(4)}  ~\equiv { \displaystyle{\sum_{1\le i<j\le n}}} s_{ij}^2\,.$

\vspace{4mm}
\noindent $\blacktriangleright$  $O(p^6)$ amplitudes:
\begin{align}
\mathcal{A}_4^{(6)}&= \frac{128}{(d-2)^4}
\frac{\beta}{f^{2d-4}}\,\big( s^3 + t^3 + u^3\big) \,,
\nonumber\\[1pt]
\mathcal{A}_5^{(6)}&= \frac{128 (d+2)}{(d-2)^5} 
\frac{\beta}{f^{5d/2-5}}\,P_5^{(6)} \,,
\nonumber\\[1pt]
\mathcal{A}_6^{(6)}&= \frac{64 (d+2)}{(d-2)^6} 
\frac{\beta}{f^{3d-6}} \,\left(4 \,d \,P_{6,A}^{(6)} + (d+2)P_{6,B}^{(6)}\right)  \,,
\nonumber \\[1pt] 
\mathcal{A}_{7}^{(6)}&= \frac{256\, d\, (d+2)}{(d-2)^7} 
\frac{\beta}{f^{7d/2-7}}\,\left((3d-2)P_{7,A}^{(6)}+ (d+2) P_{7,B}^{(6)} \right)\,,
 \nonumber\\[1pt]
\mathcal{A}_{8}^{(6)}&= \frac{256\, d\, (d+2) (5d-2)}{(d-2)^8} 
\frac{\beta}{f^{4d-8}}\,\left(2(d-2) P_{8,A}^{(6)}+(d+2)P_{8,B}^{(6)}\right)\,,
\label{ADimd_p6}
\end{align}
where  $P^{(6)}_{n,A} ~= \sum\limits_{1\leq i<j\leq n}s_{ij}^3
 ~\text{and}~P^{(6)}_{n,B}~ =  \sum\limits_{1\leq i<j<k\leq n}s_{ijk}^3\,.$

\vspace{4mm}
\noindent $\blacktriangleright$  $O(p^8)$ amplitudes:\\[2mm]
{\allowdisplaybreaks[2] \small
\begin{align}
\mathcal{A}_4^{(8)}&=
\frac{1}{f^{2 d-4}}
	\frac{32 }{(d-2)^4} 
	\,\left(9 \gamma +\tilde{\gamma }\right)\big( s^4 + t^4 + u^4\big) \,, 
\nonumber\\[2pt]
\mathcal{A}_5^{(8)}&=
\frac{1}{f^{\frac{5 d}{2}-5}}\frac{32}{(d-2)^5} \left(\left[9 (d+4) \gamma +d \tilde{\gamma }\right] \,P_{5,A}^{(8)} ~+~ 8 \tilde{\gamma }\, P_{5,B}^{(8)}\right)
\nonumber\\[2pt]
\mathcal{A}_6^{(8)}&= \frac{1}{f^{3 d-6}}\frac{16}{(d-2)^6} \left( 4\,\left[9 (d+4) (d+1) \gamma + d\,(d-1)\tilde{\gamma }\right] \,P_{6,A}^{(8)} +\left[9 (d+4)^2 \gamma +d^2 \tilde{\gamma }\right] \,P_{6,B}^{(8)}   
\right.\nonumber \\
&\quad 
\hspace{1.3cm} \left.
  +16\, \, (d+6)\,\tilde{\gamma } \,P_{6,C}^{(8)} +16\,  d  \,\tilde{\gamma }\,P_{6,D}^{(8)}\right)
\nonumber\\[2pt]
\mathcal{A}_7^{(8)}&=
\frac{1}{f^{\frac{7 d}{2}-7}} \frac{64 (d+1) }{(d-2)^7} \left( d\left[27\, (d+4) \gamma + (3 d+4) \tilde{\gamma }\right] \,P_{7,A}^{(8)} +\left[9 (d+4)^2 \gamma +d\,(d-8) \tilde{\gamma }\right] \, P_{7,B}^{(8)} \right.  \nonumber \\
&\quad
\hspace{1.3cm} \left.
  + 16(d+6)   \tilde{\gamma }\,P_{7,C}^{(8)} +16\, d\,   \tilde{\gamma }\,P_{7,D}^{(8)}\right)
\nonumber\\[2pt]
\mathcal{A}_8^{(8)}&=
\frac{1}{f^{4 d-8} } \frac{64\,(d+1)}{3 (d-2)^8}\left(2\, d\, \left[81 \left(2 d^2+7 d-4\right) \gamma +\left(18 d^2+7 d-4\right) \tilde{\gamma }\right]\,P_{8,A}^{(8)}\right.  \nonumber \\
&\quad
\hspace{1.3cm} \left.
 +\,d \,\left[81 (d+4)^2 \,\gamma +\left(9 d^2+4 d+8\right) \tilde{\gamma }\right]\,P_{8,B}^{(8)} \right.  \nonumber \\
&\quad
\hspace{1.3cm} \left.
  +\left[27\, (d+1) (d+4)^2\, \gamma +d \,\left(3 d^2-29 d-16\right) \tilde{\gamma }\right]\,P_{8,C}^{(8)} \right.  \nonumber \\
&\quad
\hspace{1.3cm} \left.
  +\,288 \,(5 d+2) \,\tilde{\gamma }\, P_{8,D}^{(8)}
  +\,24 \,d\, (5 d+2)\, \tilde{\gamma }\,P_{8,E}^{(8)} \right)\,,
\label{ADimd_p8}
\end{align}
}
where 
{\allowdisplaybreaks[2] 
\bea
\label{basisP8_5pt}
P_{5,A}^{(8)} &=& \sum\limits_{1\leq i<j\leq 5} s_{ij}^4
\,,\hspace{1.8cm}
P_{5,B}^{(8)} ~=~ s_{12}^2 s_{34}^2 + \text{perms} \,,
\\[11pt]
\nonumber
P_{6,A}^{(8)} &=& \sum\limits_{1\leq i<j\leq 6} s_{ij}^4
\,,\hspace{1.8cm}
P_{6,B}^{(8)} ~=~ \sum\limits_{1\leq i<j<k\leq 6} s_{ijk}^4\,,
\\[1pt]
P_{6,C}^{(8)} &=& s_{12}^2 s_{34}^2 + \text{perms} 
\,,\hspace{1.05cm}
P_{6,D}^{(8)} ~=~ s_{123}^2 s_{45}^2 + \text{perms} 
\label{basisP8_6pt}
\\[17pt]
\nonumber
P_{7,A}^{(8)} &=& \sum\limits_{1\leq i<j\leq 7} s_{ij}^4 
\,,\hspace{1.8cm}
P_{7,B}^{(8)} ~=~ \sum\limits_{1\leq i<j<k\leq 7} s_{ijk}^4 \,,
\\[1pt]
P_{7,C}^{(8)} &=& s_{12}^2 s_{34}^2 + \text{perms} 
\,,\hspace{1.05cm}
P_{7,D}^{(8)} ~=~ s_{123}^2 s_{456}^2 + \text{perms} \,,
\label{basisP8_7pt}
\eea
}
\bea
\nonumber
P_{8,A}^{(8)} &=& \!\!\sum\limits_{1\leq i<j\leq 8} s_{ij}^4\,,
\hspace{8mm}
P_{8,B}^{(8)} ~=~ \!\!\sum\limits_{1\leq i<j<k\leq 8} s_{ijk}^4\,,
\hspace{8mm}
P_{8,C}^{(8)} ~=~ \!\!\sum\limits_{1\leq i<j<k<l\leq 8} s_{ijkl}^4\,,
\\[2pt]
P_{8,D}^{(8)} &=& s_{12}^2 s_{34}^2 + \text{perms} \,,
\hspace{1.05cm}
P_{8,E}^{(8)} ~=~ s_{123}^2 s_{456}^2 + \text{perms} \,.
\label{basisP8_8pt}
\eea
Here ``$+ \text{perms}$'' includes all \emph{inequivalent} permutations of the external particle labels.

\setcounter{equation}{0}
\section{Free massive scalar flow: 1-loop dilaton scattering}
\label{app:oneloop}

Here we provide some practical details of the calculation of the 1-loop dilaton scattering amplitudes in the example of the free massive scalar in $d$-dimensions. 

Consider a 1-loop diagram with the $n$ external outgoing momenta $p_1$, $p_2$, $\dots$, $p_n$ in canonical order; all other diagrams of the same topology are obtained from the one with canonical ordering by simple permutations of the momentum labels in the result.  Momentum conservation is enforced as $\sum_{i=1}^n p_i^\mu =0$ with all momenta outgoing. The $n$ external $\varphi$'s connect to a $\Phi$-loop via $\Phi^2 \varphi^k$ terms generated by expanding the action \reef{SdDim} with $\Omega = f^{(d-2)/2} - \varphi$, as discussed in section \ref{ss:scalard}. We denote a canonical  diagram with $V$ vertices by $\{N_1, N_2, \dots, N_V\}$, where  $N_j$ are the number of external $\varphi$'s at the $j^{th}$ vertex and $\sum_{j=1}^{V}N_j=n$. For example, for $n=6$ two distinct box diagrams are\\[-5mm]
{\small
\bea
  \nonumber
  \includegraphics[trim = 0 0 0 -10px,clip,height=55px]{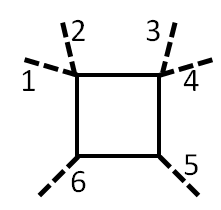}
  &~~~~&
\raisebox{2pt}{\includegraphics[trim = 0 0 0 -10px,clip,height=55px]{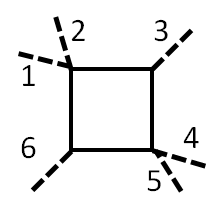}}
 \\
   \{2,2,1,1\}&~~~~& ~~ \{2,1,2,1\}
  \nonumber
\eea
}

Let $\ell$ be the loop momentum flowing into the vertex associated with $p_1$. The momentum of the $j^{th}$ internal propagator (going out of the $j^{th}$ vertex) is  $\ell - \mathcal{P}_j$, where by momentum conservation\\[-7mm]
\be
  \mathcal{P}_j 
  ~~\equiv 
  \sum\limits_{r=1}^{N_1+N_2+\ldots +N_j}p_r \,,
\hspace{2cm}
\raisebox{-1.5cm}
{\includegraphics[trim = 0 0 0 -10px,clip,height=100px]{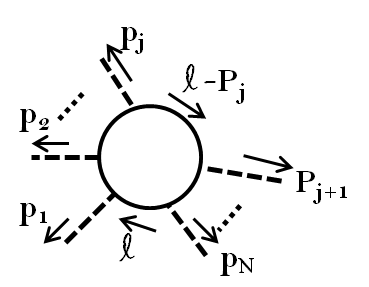}}
\ee
The expression for a canonical diagram $\{N_1,N_2,\dots ,N_V\}$ can  be written
\begin{align}
I_{\{N_1,N_2,...N_V\}} &= \frac{1}{S} \int \frac{d^d\ell}{(2\pi)^d} \prod\limits_{j=1}^{V}  V_{N_j} \frac{-i}{(\ell-\mathcal{P}_j)^2+M^2}\,,
\label{I1}
\end{align}
where $V_{N_j}$ is the vertex factor \reef{vertexrule} associated with the $j^{th}$ vertex.
The symmetry factor $S$ takes into account exchanges of identical internal propagators. 
All diagrams we consider have $S=1$ except the bubble diagrams with exactly two vertices, which have $S=2$. 

It is useful to Feynman-parameterize \reef{I1} as
\begin{align}
&I_{\{N_1,N_2,...N_V\}}  \label{I2}\\
&\quad =\frac{(-1)^{V+N}}{S}\frac{M^{2V}}{f^N} \left(\prod\limits_{n=0}^{N_j-1} \left( \tfrac{4}{d-2}-n \right)  \right)
\displaystyle\int
\frac{d^d\ell}{(2\pi)^d} 
\left(\prod\limits_{j=1}^{V}\displaystyle\int_{0}^{1} dx_j\right) \frac{\Gamma(V)~\delta\Big(\scriptstyle 1-\sum\limits_{k=1}^{V}x_k\Big)}{\left[\sum\limits_{m=1}^{V}x_m\left((\ell-\mathcal{P}_m)^2+M^2\right)\right]^V}\,.
    \nonumber
\end{align}
We are interested in the low-energy expansion of the amplitudes, so we expand the integrals \reef{I2} in the Mandelstam invariants of the external momenta. Practically this is done by shifting the loop-momentum $\ell$ such that the integrand can be expanded in powers of 
${\mathbb{P}^2}/(\ell^2+M^2)$, where 
\be 
  \mathbb{P}^2 
  \equiv 
  \bigg(\sum\limits_{m=1}^{V}x_m\mathcal{P}_m\bigg)^2 
     - \sum\limits_{m=1}^{V}x_m\mathcal{P}_m^2\,.
\ee
A little algebra shows that the $O(p^{2k})$ part of the diagram is
\begin{align}
I_{\{N_1,N_2,...N_V\}}^{O(p^{2k})} 
&=\frac{(-1)^{V+N}}{S}\frac{M^{2V}}{f^N}
\frac{\Gamma(V+k)}{k!} 
\left(\prod\limits_{n=0}^{N_j-1} \left( \tfrac{4}{d-2}-n \right) \right)\bigg( \displaystyle\int \frac{d^d\ell}{(2\pi)^d}\frac{1}{\left[\ell^2+M^2\right]^{V+k}}\bigg) 
\nonumber\\
     & \hspace{5cm} \times \,
     \!\bigg[\bigg(\prod\limits_{j=1}^{V} \int_{0}^{1} \!\!\! dx_j 
      \bigg)
      \,\left(\mathbb{P}^2\right)^k\, 
      \delta\Big(\scriptstyle 1- \sum\limits_{k=1}^{V}x_k\Big)
      \bigg]
    \,.
\end{align}
The momentum integral is finite for $V+k>d/2$ and gives (in Euclidean signature)
\begin{align}
\int_{-\infty}^{\infty} \frac{d^d\ell}{(2\pi)^d}\frac{1}{\left[\ell^2+M^2\right]^{V+k}} &= \int d\Omega_{d-1} \int_0^{\infty} \frac{d\ell}{(2\pi)^d}\frac{\ell^{d-1}}{\left[\ell^2+M^2\right]^{V+k}} \nonumber \\
&=\frac{1}{(4\pi)^{d/2}M^{2(V+k-d/2)}} 
\frac{\Gamma(V+k-d/2)}{\Gamma(V+k)}
\end{align}
So we arrive at the Mathematica-friendly expression:
\bea
&&I_{\{N_1,N_2,...N_V\}}^{O(p^{2k})} 
\\[-5mm]
 \nonumber
&&~~= \frac{(-1)^{V+N}}{S}\,
\frac{\Gamma(V+k-d/2)}{k!}
\frac{ M^{d-2k}}{f^N(4\pi)^{d/2}} 
\left(\prod\limits_{n=0}^{N_j-1} \left( \tfrac{4}{d-2}-n \right)  \right) 
      \!\bigg[\bigg(\prod\limits_{j=1}^{V} \!\!\! \int\limits_{0}^{1-\sum\limits_{q=1}^{j-1}x_q} \!\!\! dx_j 
      \bigg)
      \,\left(\mathbb{P}^2\right)^k\bigg]
    \,.
\eea

To obtain the full contribution from diagrams of a given topology $\{N_1,N_2,...N_V\}$, we must sum over inequivalent permutations of the external momenta, i.e.~over arrangements of the external momentum labels not related by cyclic permutations or reflection symmetry. The final result can be written in terms of a basis of Mandelstam polynomials which are fully symmetric in the external momenta, e.g.~the $O(p^8)$ basis of \reef{basisP8_5pt}-\reef{basisP8_8pt}.

\vspace{2mm}
\noindent {\bf Example: 4-point amplitude.}
Consider the 4-point amplitude at $\mathcal{O}(p^8)$ in $d=8$ dimensions. There are 3 types of diagrams, a bubble, a triangle and a square. The canonical diagram $\{2,1,1\}$ gives
\be
\raisebox{-8mm}
{\includegraphics[trim = 0 0 0 10px,clip,width=100px]{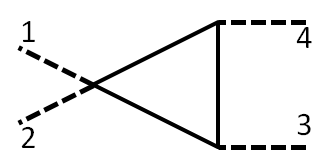}}
~=~
\frac{s^4}{612\,360\, (4\pi)^4\, f^{12}}.
\ee
There are ${4 \choose 2} = 6$ distinct permutations of the momentum labels of this diagram, and summing them gives the result in the first row of the following table:\\[1mm]

{\newcolumntype{C}{>{\centering\arraybackslash} m{0.2\textwidth} }
\begin{tabular}{| C | C | C | C |} 
\hline
\textbf{Diagram} & \textbf{Unique Permutations} & \textbf{Symmetry Factor} & \textbf{Partial Amplitude at $\mathcal{O}(p^8)$ in 8d}\\ \hline
\includegraphics[trim = 0 0 0 -10px,clip,height=30px]{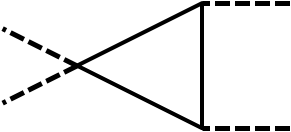} & 6 & 1 
& 
$\dfrac{s^4+t^4+u^4}{612\,360\, (4\pi)^4\, f^{12}}$
\\ \hline
\includegraphics[trim = 0 0 0 -10px,clip,height=30px]{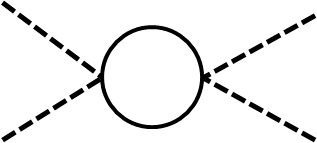} & 3 & 2 
&
$\dfrac{s^4+t^4+u^4}{382\,725\, (4\pi)^4\, f^{12}}$
\\ \hline
\includegraphics[trim = 0 0 0 -10px,clip,height=30px]{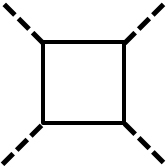} & 3 & 1 
&$\dfrac{s^4+t^4+u^4}{765\,450\, (4\pi)^4\, f^{12}}$
\\\hline
\end{tabular}
}
\vspace{8mm}

\noindent The sum of  the contributions from the three classes of diagrams in the table gives the result for the 4-point 1-loop amplitude at $\mathcal{O}(p^8)$ in $8d$, 
$
\mathcal{A}_4^{(8)}=\frac{17}{3\,061\,800 \, (4\pi)^4\, f^{12}}
$,
as also listed in \reef{Anp8inDim8-1loop}.


\end{document}